# The propagation of transient waves in two-dimensional square lattices

Nadezhda I. Aleksandrova (*nialex@misd.ru*)

Chinakal Institute of Mining, Siberian Branch of the Russian Academy of Sciences, Krasnyj pr. 54, Novosibirsk, 630091 Russia

**Abstract:** The aim of this article is to study the attenuation of transient low-frequency waves in 2D lattices in both plane and antiplane problems. The main idea of this article is that analytical solutions to problems of mechanics of discrete periodic media can be obtained by a method of asymptotic inversion of the Laplace and Fourier transforms in the vicinity of the quasi-front of infinitely long waves; moreover, in this method it is possible to take into account the contribution of short waves. Using this method, we obtain asymptotics of perturbations in lattices in plane and antiplane formulations under a local transient load. Besides, we show that equations describing 2D plane motion of a square lattice can be represented in the form of two linearly independent wave equations, each of which contains one unknown function only. By analogy with the theory of elasticity, one equation describes the propagation of shear waves in the lattice, while the other equation describes the propagation of longitudinal waves. As a result, it is shown that, in a homogeneous infinite lattice, a load can be specified in such a manner that either predominantly longitudinal or predominantly shear waves are formed. The problems under study are also solved by a finite difference method. The qualitative and quantitative correspondence of asymptotic and numerical solutions is shown.

### Highlights

- Wave equations describing the plane motion of a 2D square lattice are splitted.
- Asymptotic solutions are obtained describing the propagation of antiplane and plane perturbations under a local step load.
- Numerical examples are given to demonstrate the accuracy of asymptotic solutions.
- For problems of mechanics of discrete-periodic media, a method is proposed for asymptotic inversion of the Fourier and Laplace transforms.

# 1. Introduction

The studies [1 – 3] show that the block structure of rocks should be taken into account in mathematical models. The fundamental concept of M.A. Sadovsky [1] had a particularly great influence on the development of this approach. According to that hypothesis, the rock mass is a system of blocks, interconnected by interlayers consisting of weaker fractured rocks. The presence of interlayers with weakened mechanical properties leads to the fact that the deformation of the block rock mass both in statics and in dynamics occurs mainly due to their pliability [2, 3]. As noted in [2, 3], the block structure of the rock mass is the cause of the appearance of low-frequency waves with low propagation velocity, long wavelength, and weak damping. In the literature, such waves are often called pendulum waves, see, for example, [2 – 5]. The simplest model of a block rock medium can be obtained if the dynamic behavior of the block medium is approximately described as the movement of rigid blocks due to deformation of the interlayers, which are modeled as springs and dampers. In this case, the block medium can be represented as a periodic lattice of point-masses. Such a representation of a block medium makes it possible to describe low-frequency waves arising from impact loading fairly accurately. The comparison of the results of field and laboratory experiments and numerical calculations made according to this approach to modeling a block rock medium is given, for example, in [4] (in the one-dimensional case) and in [5] (in the 3D case). In the present article, a block medium is modeled as a 2D square lattice of masses connected by springs either in the axial directions only, or in the axial and diagonal directions.

Note that in the literature, the mass lattice model is used not only for modeling rock masses, but also for modeling crystal structures. In publications devoted to crystal lattices, the main attention is paid to the study of the pass- and stop-bands in the frequency domain for band-gap materials. There are many such publications. We refer the interested reader, for example, to [6 – 12]).

Other aspects of the study of lattices are presented in the books [13, 14] and other publications of the authors of these books, see, for example, [15, 16]. They contain analytical methods, formulations of problems and their solutions, and an overview of the results devoted to the problems of crack propagation and the effects of wave localization in inhomogeneous or disordered discrete structures and lattices.

In many publications, the behavior of lattice Green's function is studied as a function of the frequency time-harmonic excitation near the critical frequencies and the dispersion properties of Bloch–Floquet waves are analyzed, see, for example, [14], [17] and references given therein. In contrast, our article examines the behavior of the solution as a function of time under transient excitation.

The problem of propagation of transient waves in discrete-periodic media has a long history. To solve transient elastic problems, Slepyan [18] proposed an asymptotic method for joint inversion of the Laplace – Fourier transforms in the vicinity of the quasi-front of a traveling wave. Subsequently, this method was used to obtain asymptotic solutions that allow as to estimate long-wave disturbances in various one-dimensional media having either a homogeneous or periodic structure, see, for example, [4, 19, 20].

Less studied are transient wave processes in 2D and 3D discrete-periodic media. Nevertheless, some results are known for such media as well. For example, in [21 – 26], some models of periodic lattices of masses are used to describe plane or antiplane viscoelastic deformations of 2D discrete media. In [21], analytical and numerical results are presented concerning the formation of "star" waves in 2D square and triangular lattices under the action of a concentrated load with a resonant frequency. In [22], an analytical solution to the antiplane problem for a square lattice of masses (masses are connected in axial directions only) is obtained under a concentrated in space step in time load. In [23], an analytical solution to the plane problem for a square lattice of masses (masses are connected in axial and diagonal directions) is obtained under a concentrated in space step in time vertical load applied on the boundary of a half-plane filled with the lattice of masses. In [24], a numerical study in a planar setting of wave propagation is carried out in a square lattice under the action of a load of the "center of dilatation" type. Transient wave processes in 3D discrete media were studied in [5, 25, 26]. In [25], a solution to the Lamb problem for 2D and 3D lattices is given in the form of multiple integrals, which are similar to the integral representations of the Bessel functions. In [5, 26], the Lamb problem in a 3D lattice is studied numerically.

Thus, the propagation of transient waves in 2D and 3D discrete-periodic media has been studied in much less detail than in one-dimensional periodic media. In particular, for waves in 2D and 3D lattices, analytical solutions are practically unknown (we know just a few exceptions, namely, [21 – 23]). The present article partially fills this gap. Namely, we find analytical solutions to a number of problems of wave propagation in 2D square lattices under the action of local transient loads. Moreover, our numerical experiments show good agreement with the asymptotic solutions found.

**2. Splitting the equations of motion in the plane problem**

Consider a uniform 2D unbounded square lattice consisting of equal masses $M$ connected by massless elastic springs in the directions of axes $x$, $y$, and in the diagonal directions (Fig. 1). The length $l$ of the springs in the directions of axes $x$, $y$ is the same and equal to one. Note that this lattice is similar to lattices used by Born and Kármán in their study of problems in solid state physics [6] and by Jensen in his study of phonon band gaps [10].

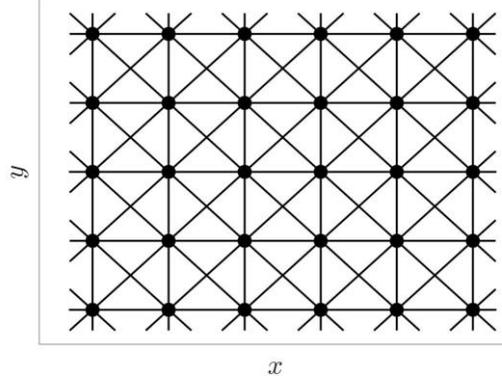

**Fig. 1.** The square lattice of the masses connected by springs in the directions of axes *x*, *y*, and in the diagonal directions.

Let *n*, *m* be the indices of the masses along axes *x*, *y*. The equations of the plane motion of the mass with indices *n*, *m* were derived in [24] and have the following form:

$$M\ddot{u}_{n,m} = k_1(u_{n+1,m} - 2u_{n,m} + u_{n-1,m}) + k_2(u_{n+1,m+1} + u_{n-1,m-1} + u_{n+1,m-1} + u_{n-1,m+1} - 4u_{n,m})/2$$
$$+ k_2(v_{n+1,m+1} + v_{n-1,m-1} - v_{n-1,m+1} - v_{n+1,m-1})/2 + P_{n,m}, \qquad (1)$$

$$M\ddot{v}_{n,m} = k_1(v_{n,m+1} - 2v_{n,m} + v_{n,m-1}) + k_2(u_{n+1,m+1} + u_{n-1,m-1} - u_{n+1,m-1} - u_{n-1,m+1})/2$$
$$+ k_2(v_{n+1,m+1} + v_{n-1,m-1} + v_{n-1,m+1} + v_{n+1,m-1} - 4v_{n,m})/2 + Q_{n,m}. \qquad (2)$$

Here $u_{n,m}$, $v_{n,m}$ are the displacements in directions *x*, *y* respectively; $P_{n,m}$, $Q_{n,m}$ are the projections of the external force on axes *x*, *y*; $k_1$ is the spring stiffness in the directions of axes *x*, *y*; $k_2$ is the spring stiffness in the diagonal directions; *M* is the mass. The parameters *M*, $k_1$, $k_2$, appearing in (1) and (2), have the same values for all points of the lattice.

From now on, we assume $k_2 = k_1/2$. As shown in [24], in this case the velocities of longitudinal and shear infinitely long waves in the lattice are independent of the direction of wave propagation. In addition, we further assume that $M = 1$, $k_1 = 1$.

In this section, we split Eqs. (1), (2) under the assumption that there are no external forces (i.e., $P_{n,m} = Q_{n,m} = 0$). For this, we introduce the operators $A_1$, $A_2$, $B_1$, $B_2$, given by the formulas:

$$A_1 f_{n,m} = f_{n+1,m} + f_{n-1,m}, \quad A_2 f_{n,m} = f_{n,m+1} + f_{n,m-1}, \quad B_1 f_{n,m} = f_{n+1,m} - f_{n-1,m}, \quad B_2 f_{n,m} = f_{n,m+1} - f_{n,m-1}. \qquad (3)$$

Using (3), we obtain:

$$f_{n+1,m} - 2f_{n,m} + f_{n-1,m} = (A_1 - 2)f_{n,m}, \qquad (4)$$

$$\frac{1}{4}\left(f_{n+1,m+1}+f_{n-1,m-1}+f_{n+1,m-1}+f_{n-1,m+1}-4f_{n,m}\right)=\left(\frac{1}{4}A_1A_2-1\right)f_{n,m}, \tag{5}$$

$$\frac{1}{4}\left(f_{n+1,m+1}+f_{n-1,m-1}-f_{n+1,m-1}-f_{n-1,m+1}\right)=\frac{1}{4}B_1B_2 f_{n,m}. \tag{6}$$

Let us introduce also the following operator:

$$Df_{n,m}=\ddot{f}_{n,m}. \tag{7}$$

Using (4)–(7), we rewrite (1), (2) as follows:

$$\left(D-A_1+3-\frac{1}{4}A_1A_2\right)u_{n,m}-\frac{1}{4}B_1B_2 v_{n,m}=0, \tag{8}$$

$$\left(D-A_2+3-\frac{1}{4}A_1A_2\right)v_{n,m}-\frac{1}{4}B_1B_2 u_{n,m}=0. \tag{9}$$

Apply the operator $\left(D-A_2+3-A_1A_2/4\right)$ to (8) and the operator $B_1B_2/4$ to (9). Add the results together. Since any two of the operators $A_1$, $A_2$, $B_1$, $B_2$ commute with each other, we obtain:

$$\left[\left(D-A_1+3-\frac{1}{4}A_1A_2\right)\left(D-A_2+3-\frac{1}{4}A_1A_2\right)-\frac{1}{16}B_1^2B_2^2\right]u_{n,m}=0. \tag{10}$$

Equation (10) can be rewritten as follows:

$$(D-D_-)(D-D_+)u_{n,m}=0,$$

where the operators $D_-$, $D_+$ are defined by the formulas

$$D_- f_{n,m}=\frac{1}{2}(A_1+A_2-4)f_{n,m}, \qquad D_+ f_{n,m}=\frac{1}{2}\left[(A_1+1)(A_2+1)-9\right]f_{n,m}.$$

Similarly, we obtain the following equation for $v_{n,m}$:

$$(D-D_-)(D-D_+)v_{n,m}=0.$$

Let us consider functions $\varphi_{n,m}$ and $\psi_{n,m}$ such that

$$(D-D_+)\varphi_{n,m}=0, \quad (D-D_-)\psi_{n,m}=0. \tag{11}$$

Note that every $\varphi_{n,m}$ satisfying $(D-D_+)\varphi_{n,m}=0$ is a solution to $(D-D_-)(D-D_+)\varphi_{n,m}=0$. Similarly, every $\psi_{n,m}$ satisfying $(D-D_-)\psi_{n,m}=0$ is a solution to $(D-D_-)(D-D_+)\psi_{n,m}=0$. For a similar splitting of some other equations, describing the elastic motion of a discrete medium the reader is referred to [25].

Passing from the operator form of Eqs. (11) to the differential-difference form, we find that the system of Eqs. (1), (2) with respect to two unknown functions $u_{n,m}$ and $v_{n,m}$ splits into two linearly independent scalar wave equations, each with respect to one unknown function $\varphi_{n,m}$ or $\psi_{n,m}$:

$$\ddot{\varphi}_{n,m} = \frac{1}{2}\left(\varphi_{n+1,m+1} + \varphi_{n-1,m-1} + \varphi_{n+1,m-1} + \varphi_{n-1,m+1} + \varphi_{n+1,m} + \varphi_{n-1,m} + \varphi_{n,m-1} + \varphi_{n,m+1} - 8\varphi_{n,m}\right), \quad (12)$$

$$\ddot{\psi}_{n,m} = \frac{1}{2}\left(\psi_{n+1,m} + \psi_{n-1,m} + \psi_{n,m-1} + \psi_{n,m+1} - 4\psi_{n,m}\right). \quad (13)$$

Let us investigate the dispersion properties of (12), (13). Substituting $\psi_{n,m} = \Psi e^{i(\omega_- t + q_x n + q_y m)}$ and $\varphi_{n,m} = \Phi e^{i(\omega_+ t + q_x n + q_y m)}$ into (12), (13), we obtain the following dispersion equations for Bloch–Floquet waves in the square lattice:

$$\omega_+^2 = 4 - \cos q_x - \cos q_y - 2\cos q_x \cos q_y, \quad (14)$$

$$\omega_-^2 = 2 - \cos q_x - \cos q_y. \quad (15)$$

Here $q_x$, $q_y$ are the projections of wave vector Bloch $\boldsymbol{q}$ on axes $x$, $y$, respectively. Note that (14), (15) coincide with the dispersion equations obtained in [24] for Eqs. (1), (2) in the case $k_2 = k_1/2$. Passing to the limit in (14), (15) as $q_y \to 0$, $q_x = \pi$ or $q_x \to 0$, $q_y = \pi$, we obtain:

$$\omega_+ = \sqrt{3/2}, \quad \omega_- = 1/\sqrt{2}.$$

From (14), (15) we determine the moduli of phase velocity vector $c_{ph}$ and group velocity vector $c_{gr}$ of infinitely long waves ($q_x, q_y \to 0$) in the lattices (we mean that frequencies $\omega_+$, $\omega_-$ correspond to phase velocity vectors $c_{ph}^+$, $c_{ph}^-$ and group velocity vectors $c_{gr}^+$, $c_{gr}^-$, respectively):

$$\left|c_{ph}^+\right| = \left|c_{gr}^+\right| = \sqrt{3/2}, \quad \left|c_{ph}^-\right| = \left|c_{gr}^-\right| = 1/\sqrt{2}.$$

Let us introduce the notation

$$c_1 = \sqrt{3/2}, \quad c_2 = 1/\sqrt{2}, \quad \omega_1 = \sqrt{3/2}, \quad \omega_2 = 1/\sqrt{2}. \quad (16)$$

As shown in [24], velocity $c_1$ ($c_2$) is equal to the velocity of longitudinal (respectively, shear) infinitely long waves in the lattice; frequency $\omega_1$ ($\omega_2$) is equal to the frequency of short waves in a longitudinal (respectively, shear) wave. Thus, (12) describes the propagation of longitudinal waves in the lattice shown in Fig. 1, while (13) describes the propagation of shear waves in a lattice in which the masses are connected in the directions of axes $x$, $y$ only, as shown in Fig. 2.

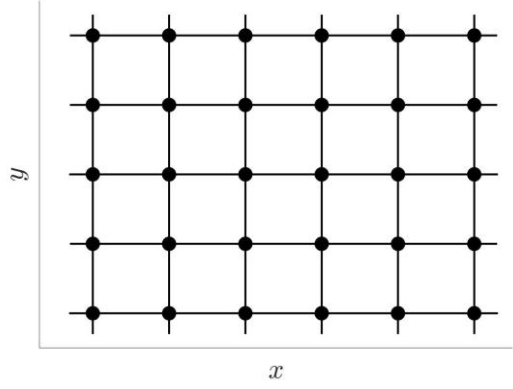

**Fig. 2.** The square lattice of the masses connected by springs in the directions of axes *x*, *y* only.

### 3. Antiplane problem. Propagation of longitudinal waves

The equations of the plane problem (1), (2) split into two scalar wave equations. In Sections 3–8, we study antiplane problems for each of these equations. There are three reasons for this: (a) each problem is of independent interest; (b) the study of these problems allows us to reveal the features of wave processes in the plane problem because, from the mathematical point of view, the wave nature of plane and antiplane processes is the same; (c) the method for solving the antiplane problem will be used to solve the plane problem in Sections 9, 10.

In this Section, we study the antiplane problem for the lattice shown in Fig. 1. Let a concentrated step load $Q$ orthogonal to the plane of the lattice acts at the point with coordinates $n=0$, $m=0$, i.e., let $Q = H(t)\delta_{0n}\delta_{0m}$, where $H$ is the Heaviside unit step function, $t$ is time, $\delta_{0n}$ is the Kronecker delta. Then the equations of the motion of the masses of the lattice are of the form:

$$\ddot{\varphi}_{n,m} = \frac{1}{2}\left(\varphi_{n+1,m+1} + \varphi_{n-1,m-1} + \varphi_{n+1,m-1} + \varphi_{n-1,m+1} + \varphi_{n+1,m} + \varphi_{n-1,m} + \varphi_{n,m-1} + \varphi_{n,m+1} - 8\varphi_{n,m}\right) + Q. \qquad (17)$$

Here $\varphi_{n,m}$ is the displacement of the lattice mass with coordinates *n*, *m* in the direction orthogonal to the lattice plane. The initial conditions for (17) are supposed to be zero.

In order to construct an analytical solution, we apply to (17) the Laplace transform with respect to time and the discrete Fourier transforms with respect to coordinates *n*, *m*:

$$f^L(p) = \int_0^\infty f(t)e^{-pt}dt, \quad f^{F_n}(q_x) = \sum_{n=-\infty}^{n=\infty} f_n e^{iq_x n}, \quad f^{F_m}(q_y) = \sum_{m=-\infty}^{m=\infty} f_m e^{iq_y m},$$

$$f(t) = \frac{1}{2\pi i}\int_{\alpha-i\infty}^{\alpha+i\infty} f^L(p)e^{pt}dp, \quad f_n(t) = \frac{1}{2\pi i}\int_{-\pi}^{\pi} f^{F_n}(q_x)e^{-iq_x n}dq_x, \quad f_m(t) = \frac{1}{2\pi i}\int_{-\pi}^{\pi} f^{F_m}(q_y)e^{-iq_y m}dq_y.$$

Here superscript $L$ corresponds to the Laplace transform in time with parameter $p$, while superscripts $F_n$, $F_m$ correspond to the discrete Fourier transforms with respect to coordinates $n$, $m$ with parameters $q_x$, $q_y$, respectively.

The Laplace–Fourier transform of $\varphi$ is as follows:

$$\varphi^{LF_nF_m}(p,q_x,q_y) = \frac{1}{pD_1(p,q_x,q_y)}, \qquad (18)$$

where

$$D_1(p,q_x,q_y) = B_1\left[1-(2\cos q_x+1)B_1^{-1}\cos q_y\right], \quad B_1 = p^2 - \cos q_x + 4. \qquad (19)$$

Let us list the asymptotic formulas derived in Appendix A (as $t\to\infty$) for displacement $\varphi_{n,0}$, velocity $\dot\varphi_{n,0}$ and acceleration $\ddot\varphi_{n,0}$ of the lattice masses in the vicinity of the longitudinal wave front:

$$\varphi_{n,0}(t) \sim \frac{1}{2\pi\omega_1^2}\begin{cases} H(\omega_1 t - n)\ln\left(\dfrac{\omega_1 t}{n}+\sqrt{\dfrac{\omega_1^2 t^2}{n^2}-1}\right), & n\neq 0, \\[2ex] \ln(4\sqrt{6}\omega_1 t)+\gamma, & n=0, \end{cases} \qquad (20)$$

$$\dot\varphi_{n,0}(t) \sim \frac{H(\omega_1 t - n)}{2\pi\omega_1\sqrt{\omega_1^2 t^2 - n^2}} \quad \text{for } n\neq 0, \qquad (21)$$

$$\dot\varphi_{n,0}(t) \sim \frac{1}{2\omega_1}\left[\frac{\text{Ai}(\kappa_{n,1})}{(\omega_1 t/2)^{1/3}}\right]^2 \quad \text{as } n\to\infty, \qquad (22)$$

$$\ddot\varphi_{n,0}(t) \sim -\frac{2\text{Ai}(\kappa_{n,1})\text{Ai}'(\kappa_{n,1})}{\omega_1 t} \quad \text{as } n\to\infty, \qquad (23)$$

where $\text{Ai}(\kappa)$ is the Airy function and

$$\kappa_{n,1} = \frac{n-\omega_1 t}{(\omega_1 t/2)^{1/3}}. \qquad (24)$$

Asymptotics of velocity (22) and acceleration (23), calculated for $t\to\infty$, take into account both long-wave perturbations ($q_x$, $q_y \to 0$) and short-wave perturbations ($q_x = 0$, $q_y = \pi$ or $q_x = \pi$, $q_y = 0$) propagating in the axial directions. Asymptotics of displacements (20) and velocities (21) take into account only the contribution of long-wave perturbations and do not take into account the contribution of short-wave perturbations.

In the case $k_1 = 2k_2$, infinitely long waves propagate in the square lattice with the same velocities in all directions. Hense, in this case, the low-frequency wave process is symmetric with respect to the point of load application. Therefore, solutions (20) – (23) can be rewritten in terms of the radial coordinate $r = \sqrt{n^2 + m^2}$:

$$\varphi_{n,m}(t) \sim \frac{1}{2\pi\omega_1^2} \begin{cases} H(\omega_1 t - r)\ln\left(\frac{\omega_1 t}{r} + \sqrt{\frac{\omega_1^2 t^2}{r^2} - 1}\right), & r \neq 0, \\ \ln\left(4\sqrt{6}\omega_1 t\right) + \gamma, & r = 0, \end{cases} \quad (25)$$

$$\dot{\varphi}_{n,m}(t) \sim \frac{H(\omega_1 t - r)}{2\pi\omega_1\sqrt{\omega_1^2 t^2 - r^2}} \quad \text{for } r \neq 0, \quad (26)$$

$$\dot{\varphi}_{n,m}(t) \sim \frac{1}{2\omega_1}\left[\frac{\text{Ai}(\kappa_{r,1})}{(\omega_1 t/2)^{1/3}}\right]^2 \quad \text{as } r \to \infty, \quad (27)$$

$$\ddot{\varphi}_{n,m}(t) \sim -\frac{2\text{Ai}(\kappa_{r,1})\text{Ai}'(\kappa_{r,1})}{\omega_1 t} \quad \text{as } r \to \infty. \quad (28)$$

**4. Analytical solution of the antiplane problem for a homogeneous elastic medium**

In this Section, we show that, for $r \neq 0$, solution (25), (26) of the antiplane problem for a lattice coincides with the solution of the antiplane problem for a homogeneous elastic medium, i.e., show that infinitely long waves propagate in the lattice as in a homogeneous elastic medium.

Recall that $l$ is the length of the springs. Replacing equation (17) by its differential approximation and passing to the limit as $l \to 0$, we obtain the following equation describing the antiplane motion of a homogeneous elastic medium:

$$\frac{\ddot{u}}{c^2} = \frac{\partial^2 u}{\partial x^2} + \frac{\partial^2 u}{\partial y^2} + Q(t)\delta(x)\delta(y). \quad (29)$$

Here $u$ is the displacement of the elastic medium in the direction orthogonal to plane $x$, $y$; $\delta(x)$ is the Dirac function; $Q(t) = H(t)$ is a step load; $c$ is the velocity of propagation of perturbations in the medium.

Applying to (29) the Laplace transform with respect to time and the Fourier transforms with respect to coordinates $x$, $y$, we obtain:

$$\dot{u}^{LF_xF_y}(p, q_x, q_y) = \frac{1}{p^2/c^2 + q_x^2 + q_y^2}.$$

Here superscripts $F_x$, $F_y$ correspond to the Fourier transforms with respect to coordinates $x$, $y$ with parameters $q_x$, $q_y$, respectively. Successively applying the inverse Fourier and Laplace transforms, we get the exact solution to (29) for displacements and their velocities:

$$u(r,t) = \frac{1}{2\pi} H(z-1) \ln\left(z + \sqrt{z^2-1}\right), \quad z = \frac{ct}{r}, \quad r = \sqrt{x^2 + y^2} \neq 0, \tag{30}$$

$$\dot{u}(r,t) = \frac{c}{2\pi} \frac{H(ct-r)}{\sqrt{c^2 t^2 - r^2}}, \quad r \neq 0. \tag{31}$$

Solution (30) coincides with the solution obtained in [18] (see formula (18.34)).

For the convenience of comparing the solution for a homogeneous elastic medium with the above obtained solution for a lattice, we substitute $c = c_1$ into (30), (31). Then the comparison of (25) and (30) shows that, for displacements, the exact solution for a homogeneous elastic medium coincides with the long-wave asymptotic solution for a lattice for $r \neq 0$. Similarly, for velocities of masses, exact solution (31) for a homogeneous elastic medium coincides with the long-wave asymptotic solution (26) for a lattice for $r \neq 0$.

### 5. Comparison of finite-difference and analytical solutions for Eq. (17)

In order to determine the limits of applicability of asymptotic solutions (25)–(28), we solve Eq. (17) also by the method of finite differences according to an explicit scheme. For this, we approximate the second time derivative in (17) by the central difference:

$$\ddot{f}_{n,m}(t) \approx \frac{f_{n,m}^{k+1} - 2 f_{n,m}^k + f_{n,m}^{k-1}}{\tau^2}, \quad k \geq 0. \tag{32}$$

Here $t = k\tau$, $\tau$ is the time step of the difference mesh, $k$ is the number of the layer in time in the difference scheme.

Figs. 3, 4 and 5 show the numerical and analytical solutions. The thin curves correspond to finite-difference solutions with time step $\tau = 0.07$, while the thick curves (solid and dashed) correspond to analytical solutions (25)–(28). As can be seen in Fig. 3, over the entire interaction interval, asymptotics (25) describes with high accuracy the low-frequency part of the displacement obtained numerically.

Fig. 4 shows that analytical solution (27) for the velocities of the masses of the lattice, in which the contribution of short-wave perturbations is taken into account, oscillates around solution (26), in which the contribution of short-wave perturbations is not taken into account. As can be seen in this figure, asymptotics (27) on the diagonal of loading ($n = m$) practically coincides with the finite-difference solution before the arrival of short-wave perturbations, while on the axis ($m = 0$) asymptotics (27) is close to the numerical solution only in the vicinity of the quasi-front of the longitudinal wave ($n = c_1 t$). Fig. 5 shows that the same conclusion is valid for the asymptotic behavior of the acceleration of the masses (28).

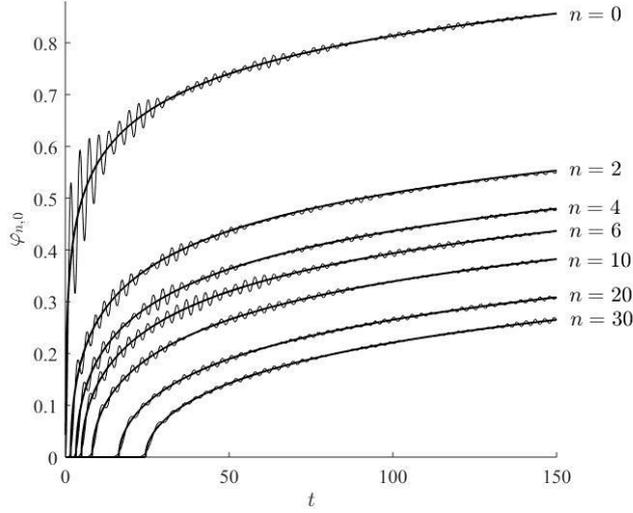

**Fig. 3.** Time dependences of displacements $\varphi_{n,0}$ of masses for $n = 0, 2, 4, 6, 10, 20, 30$. Thin curves correspond to finite-difference solutions; thick curves correspond to analytical solution (25).

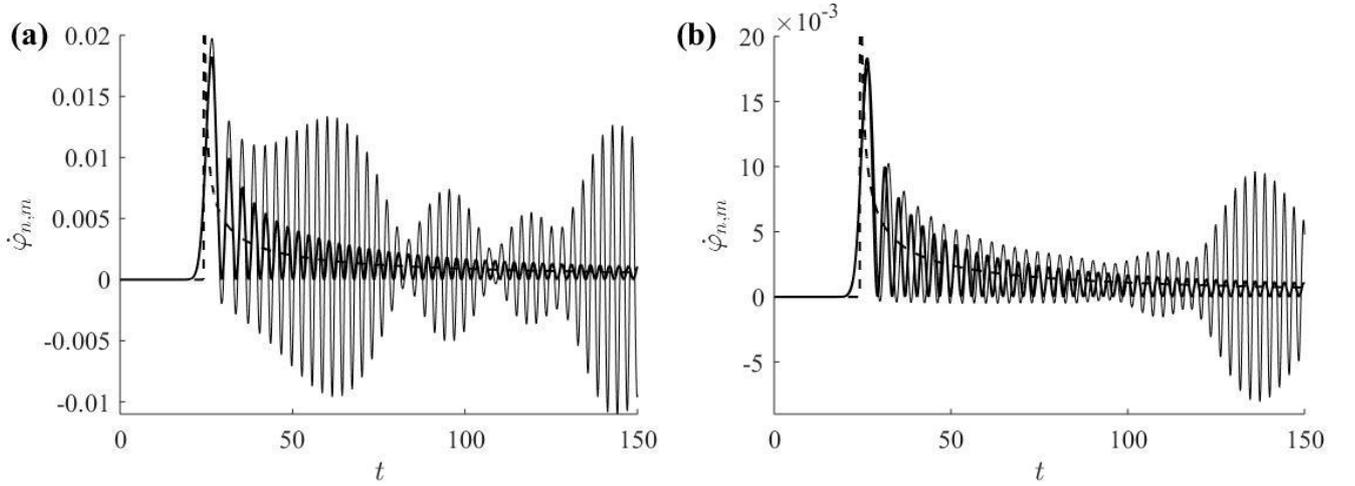

**Fig. 4.** Time dependences of velocities $\dot\varphi_{n,m}$ of masses: (a) $n = 30, m = 0$; (b) $n = m = 21$. Thin curves correspond to finite-difference solutions; thick solid curves correspond to analytical solution (27); thick dashed curves correspond to analytical solution (26).

### 6. Antiplane problem. Propagation of shear waves

In this Section, we study the antiplane problem for the lattice shown in Fig. 2. Let a concentrated step load $Q$ orthogonal to the plane of the lattice acts at the point with coordinates $n = 0$, $m = 0$, i.e., let $Q = H(t)\delta_{0n}\delta_{0m}$. Then the equations of the motion of the masses of the lattice are of the form:

$$\ddot\psi_{n,m} = \frac{1}{2}\left(\psi_{n+1,m} + \psi_{n-1,m} + \psi_{n,m-1} + \psi_{n,m+1} - 4\psi_{n,m}\right) + Q. \qquad (33)$$

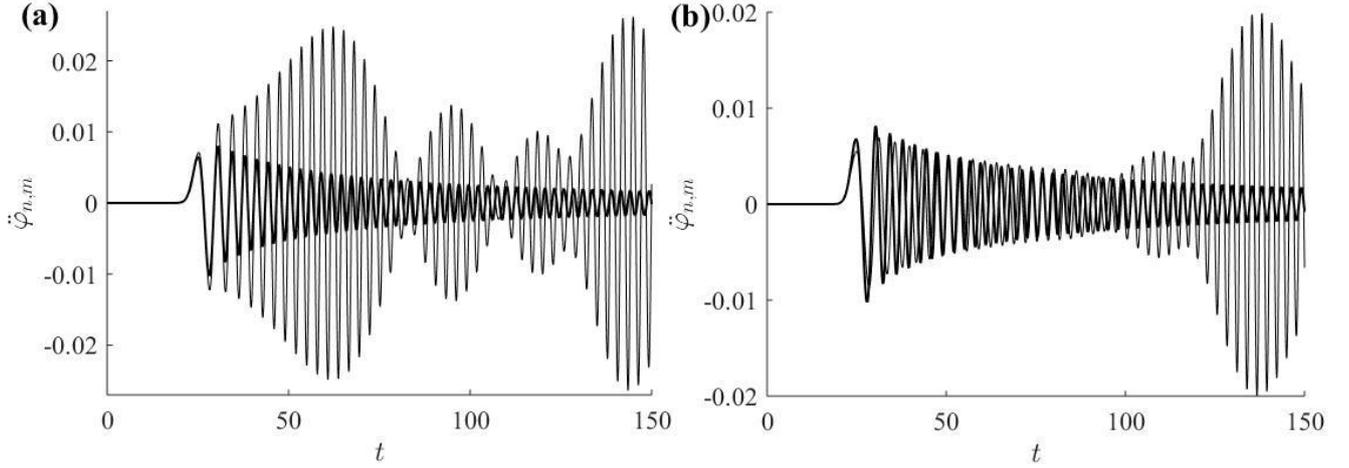

**Fig. 5.** Time dependences of accelerations $\ddot\varphi_{n,m}$ of masses: (a) $n=30, m=0$; (b) $n=m=21$. Thin curves correspond to finite-difference solutions; thick curves correspond to analytical solution (28).

Here $\psi_{n,m}$ is the displacement of the lattice mass with coordinates $n$, $m$ in the direction orthogonal to the lattice plane. The derivation of (33) is similar to that presented in [24] and is not given here. The initial conditions for (33) are supposed to be zero.

In order to construct an analytical solution, we apply to (33) the Laplace transform with respect to time and the discrete Fourier transforms with respect to coordinates $n$, $m$.

The Laplace–Fourier transform of $\psi$ is as follows:

$$\psi^{LF_nF_m}(p,q_x,q_y) = \frac{1}{pD_2(p,q_x,q_y)}, \tag{34}$$

where

$$D_2(p,q_x,q_y) = B_2\left(1 - B_2^{-1}\cos q_y\right) \quad \text{and} \quad B_2 = p^2 + 2 - \cos q_x. \tag{35}$$

From (34), one can obtain asymptotic expressions for function $\psi_{n,0}$ and its derivatives with respect to time, which describe the behavior of perturbations in the vicinity of the quasi-front of the shear wave ($c_2 t = n$). This can be done using transformations similar to those detailed in Appendix A for longitudinal wave. Omitting details, we indicate these asymptotic formulas:

$$\psi_{n,0}(t) \sim \frac{1}{2\pi\omega_2^2}\begin{cases} H(\omega_2 t - n)\ln\left[\dfrac{\omega_2 t}{n} + \sqrt{\dfrac{\omega_2^2 t^2}{n^2} - 1}\right], & n \neq 0, \\[2mm] \ln\left(4\sqrt{2}\omega_2 t\right) + \gamma, & n = 0, \end{cases} \tag{36}$$

$$\dot\psi_{n,0}(t) \sim \frac{H(\omega_2 t - n)}{2\pi\omega_2\sqrt{\omega_2^2 t^2 - n^2}} \quad \text{for} \quad n \neq 0, \tag{37}$$

$$\dot{\psi}_{n,0}(t) \sim \frac{1}{2\omega_2} J_n^2(\omega_2 t) \quad \text{as} \quad n \to \infty, \tag{38}$$

$$\ddot{\psi}_{n,0} \sim J_n(\omega_2 t) J_n'(\omega_2 t) \quad \text{as} \quad n \to \infty. \tag{39}$$

Here $J_n$ is the Bessel function of the first kind of integer order $n$ [28].

Using (A4) and (A7), we obtain from (38), (39) as $t \to \infty$:

$$\dot{\psi}_{n,0}(t) \sim \frac{1}{2\omega_2} \left[ \frac{\text{Ai}(\kappa_{n,2})}{(\omega_2 t/2)^{1/3}} \right]^2 \quad \text{as} \quad n \to \infty, \tag{40}$$

$$\ddot{\psi}_{n,0}(t) \sim -\frac{2\text{Ai}(\kappa_{n,2})\text{Ai}'(\kappa_{n,2})}{\omega_2 t} \quad \text{as} \quad n \to \infty, \tag{41}$$

where

$$\kappa_{n,2} = \frac{n - \omega_2 t}{(\omega_2 t/2)^{1/3}}. \tag{42}$$

Just as in Section 3, we rewrite asymptotic solutions (36), (37), (40), and (41), valid for $t \to \infty$, in terms of the radial coordinate $r = \sqrt{n^2 + m^2}$:

$$\psi_{n,m}(t) \sim \frac{1}{2\pi\omega_2^2} \begin{cases} H(\omega_2 t - r) \ln\left[\dfrac{\omega_2 t}{r} + \sqrt{\dfrac{\omega_2^2 t^2}{r^2} - 1}\right], & r \neq 0, \\ \ln\left(4\sqrt{2}\omega_2 t\right) + \gamma, & r = 0, \end{cases} \tag{43}$$

$$\dot{\psi}_{n,m}(t) \sim \frac{H(\omega_2 t - r)}{2\pi\omega_2 \sqrt{\omega_2^2 t^2 - r^2}} \quad \text{for} \quad r \neq 0, \tag{44}$$

$$\dot{\psi}_{n,m}(t) \sim \frac{1}{2\omega_2} \left[ \frac{\text{Ai}(\kappa_{r,2})}{(\omega_2 t/2)^{1/3}} \right]^2 \quad \text{as} \quad r \to \infty, \tag{45}$$

$$\ddot{\psi}_{n,m}(t) \sim -\frac{2\text{Ai}(\kappa_{r,2})\text{Ai}'(\kappa_{r,2})}{\omega_2 t} \quad \text{as} \quad r \to \infty. \tag{46}$$

The comparison of solutions for longitudinal wave (25)–(28) and for shear wave (43)–(46) obtained for a concentrated step antiplane load shows their qualitative agreement with each other. Under such a load, the following effects are observed in the lattice as $t \to \infty$ (or, which is the same, $r \to \infty$): (a) the amplitude of displacements is proportional to $\ln(t/r)$ for $r \neq 0$; (b) the maximum amplitude of the velocities of masses decrease as $t^{-2/3}$ ($r^{-2/3}$); (c) the maximum amplitude of the accelerations of masses decrease as $t^{-1}$ ($r^{-1}$); (d) the quasi-front zone, i.e., zone, where perturbations vary from zero to maximum, expands as $t^{1/3}$ ($r^{1/3}$). Quantitatively, the solutions for shear and longitudinal waves differ from each other in the velocity of the propagation of the quasi-front, in the frequency of oscillations behind the quasi-front, and in the amplitude of perturbations.

### 7. Comparison of finite-difference and analytical solutions for Eq. (33)

Figs. 6–8 show the displacements, velocities, and accelerations of the masses as functions of time under a concentrated step load, obtained numerically and analytically for equation (33). The thin curves correspond to finite-difference solutions obtained using approximation (32) with time step $\tau = 0.07$, while the thick curves (solid and dashed) correspond to analytical solutions (43)–(46).

As can be seen in Fig. 6, over the entire interaction interval, asymptotics (43) describes with high accuracy the low-frequency part of the displacement of masses obtained numerically. As can be seen in Fig. 7a and b, the asymptotics of the velocities of masses (45) coincides with the results of numerical calculations on axis $m = 0$ better than on diagonal $n = m$. A similar conclusion is valid for the accelerations of masses (see Fig. 8a and b). Note that, for a longitudinal wave, the opposite situation takes place: the asymptotics and numerical calculation coincide better on the diagonal than on the axis (see Figs. 4 and 5). Figs. 3–8 show that the numerical and asymptotic solutions for a lattice begin to coincide with each other at a finite distance from the loading point and at a finite time from the moment when the load is applied.

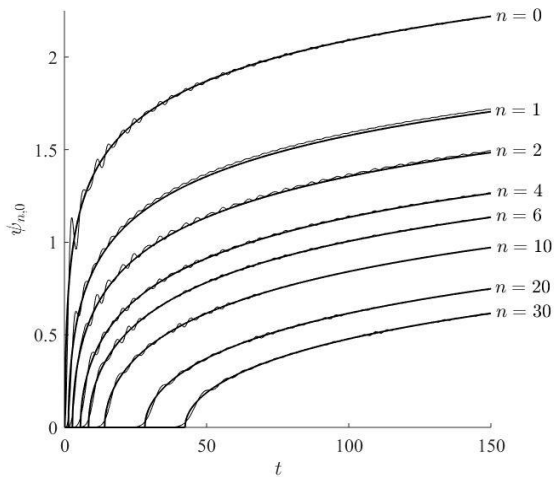

**Fig. 6.** Time dependences of the displacements of masses $\psi_{n,0}$ for $n = 0, 1, 2, 4, 6, 10, 20, 30$. Thin curves correspond to finite-difference solutions; thick curves correspond to analytical solution (43).

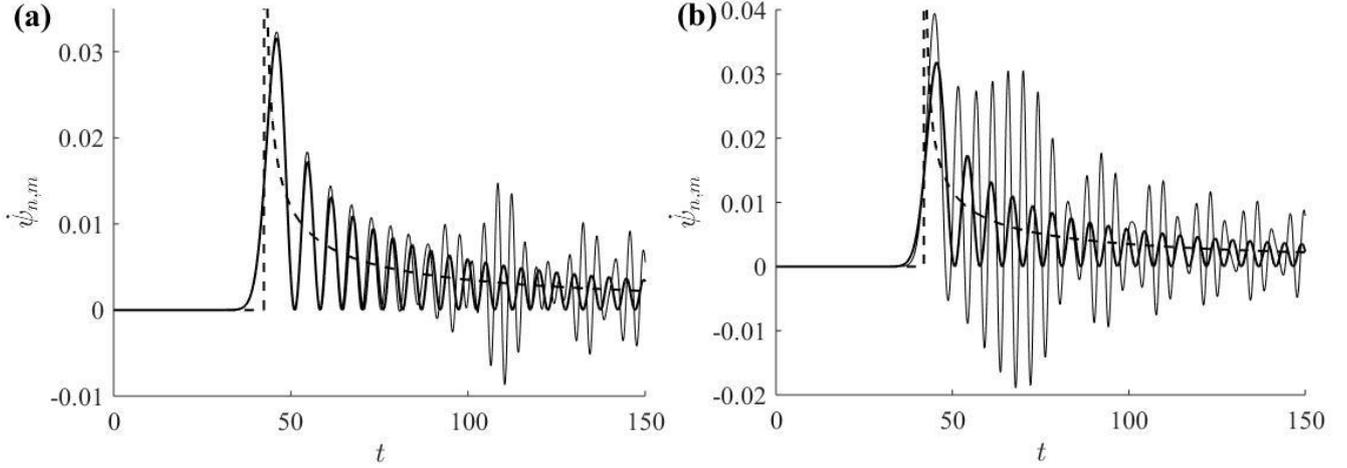

**Fig. 7.** Time dependences of velocities $\dot\psi_{n,m}$ of the masses: (a) $n=30, m=0$; (b) $n=m=21$. Thin curves correspond to the finite-difference solutions, thick solid curves correspond to analytical solution (45) and thick dashed curves correspond to analytical solution (44).

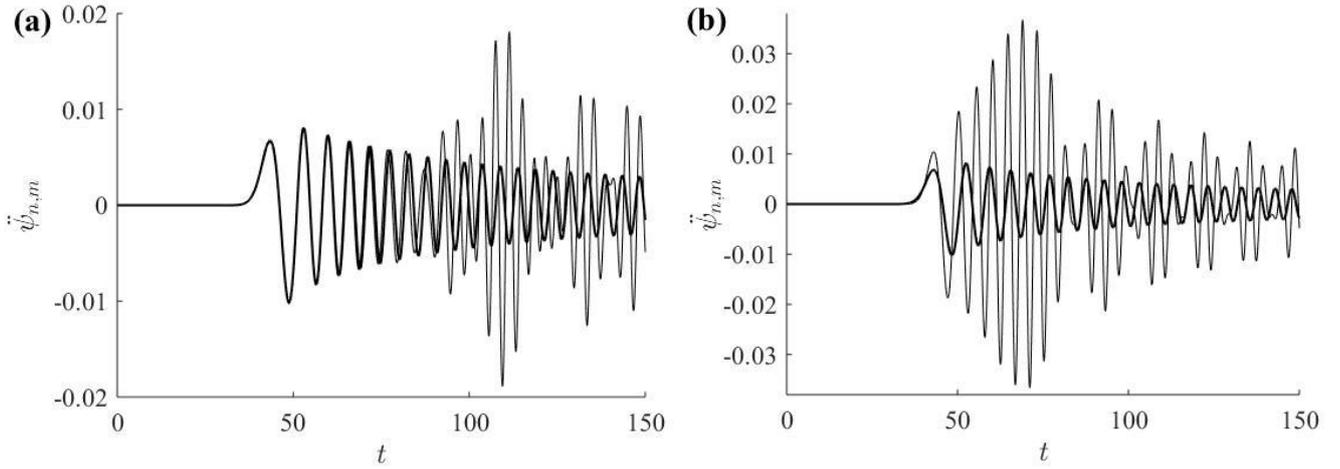

**Fig. 8.** Time dependences of accelerations $\ddot\psi_{n,m}$ of the masses: (a) $n=30, m=0$; (b) $n=m=21$. Thin curves correspond to the finite-difference solutions, thick solid curves correspond to analytical solution (46).

### 8. Propagation of resonant shear waves in the lattice

Let us return to the antiplane problem for a lattice described by Eq. (33) for displacement $\psi_{n,m}$ of the lattice mass with coordinates $n$, $m$. In Section 8, we assume that a monochromatic load $Q(t)$ of frequency $\omega_*$ is applied at the origin (0,0), i.e., we put $Q(t) = \sin(\omega_* t) H(t) \delta_{0n} \delta_{0m}$. Since $\sin(\omega_* t) = \operatorname{Im} \exp(i\omega_* t)$, we first find an analytical solution to Eq. (33) for the load $P(t) = \exp(i\omega_* t) H(t) \delta_{0n} \delta_{0m}$.

The Laplace–Fourier transform of $\psi_{n,m}$ is as follows:

$$\psi^{LF_nF_m}(p,q_x,q_y) = \frac{1}{(p-i\omega_*)D_2(p,q_x,q_y)},$$

where $D_2$ is defined by (35).

Let us find an asymptotic solution to Eq. (33) in the case of the above-specified monochromatic load as $t \to \infty$. As shown in [21], there are two resonance frequencies for the lattice of masses under consideration: $\omega_* = 2\omega_2$ ($q_y = \pm\pi \pm q_x$) and $\omega_* = 2\sqrt{2}\omega_2$ ($q_y = \pm q_x = \pm\pi$).

Put $\omega_* = 2\omega_2$ and $p = s + i\omega_*$. Then

$$\psi_n^{LF_m}(s,q_x) \sim \frac{i(-1)^{n+1} e^{i|q_x n|}}{2s\sqrt{\sin^2 q_x + 4is\omega_2^{-1}\cos q_x}} \quad \text{as} \quad s \to 0^+.$$

Applying the inverse Fourier transform with respect to $q_x$ to both sides of the last formula, we obtain:

$$\psi_{n,m}^L(s) \sim \frac{i(-1)^{n+1}}{2\pi s}\int_0^\pi \frac{\cos(q_x|m|)\exp(iq_x|n|)}{\sqrt{\sin^2 q_x + 4is\omega_2^{-1}\cos q_x}}dq_x \quad \text{as} \quad s \to 0^+. \tag{47}$$

The asymptotics of functions $\psi_{n,m}^L(s)$ as $s \to 0^+$ (i.e., the asymptotics of integrals (47)) and the asymptotics of the originals $\psi_{n,m}(t)$ of these functions as $t \to \infty$ are as follows:

a) if $|m|+|n|$ is even, then

$$\psi_{0,0}^L(s) \sim -\frac{1}{2\pi s}\ln\frac{4\omega_2}{s}, \quad \psi_{0,0}(t) \sim -\frac{1}{2\pi}\big[\ln(4\omega_2 t)+\gamma\big], \tag{48}$$

$$\psi_{n,m}^L(s) \sim \frac{(-1)^n}{2\pi s}\left(\ln\frac{s|n|}{\omega_2}+\gamma\right), \quad \psi_{n,m}(t) \sim -\frac{(-1)^n}{2\pi}\ln\frac{\omega_2 t}{|n|}, \quad |m|=|n|\neq 0, \tag{49}$$

$$\psi_{n,m}^L(s) \sim \frac{(-1)^n}{2\pi s}\left(\ln\frac{s|m^2-n^2|}{\omega_2}+2\gamma\right), \quad \psi_{n,m}(t) \sim -\frac{(-1)^n}{2\pi}\left(\ln\frac{\omega_2 t}{|m^2-n^2|}-\gamma\right), \quad |m|\neq|n|, \tag{50}$$

b) if $|m|+|n|$ is odd, then

$$\psi_{n,m}^L(s) \sim \frac{(-1)^{\nu+1}}{4s}, \quad \psi_{n,m}(t) \sim \frac{(-1)^{\nu+1}}{4}, \quad \nu = \max(|m|,|n|). \tag{51}$$

We obtain (48)–(51) by a method which was proposed in [33] for finding the asymptotics of integrals (47) in the particular case $n=m=0$. This particular case is also published as a problem in The American Mathematical Monthly in [34]. It is known that the Monthly editors select the best solution among those sent in by the readers for every problem and publish it. In the case of problem [34], the best solution was published in [35]. It was obtained by a method different from that used in [33]. Note that the same results are obtained both in [33] and [35]. The coincidence of the results obtained in [33] and [35] convinces us of the correctness of using the method proposed in [33] and, hence, formulas (48)–(51) obtained by that method. In addition, we checked the correctness of asymptotics (48)–(51) by comparing them with numerical solutions to Eq. (33) for arbitrary $m$ and $n$.

The problem of finding the asymptotics of integrals (47) as $s \to 0^+$ was also solved in [21]. However, there are inaccuracies in some formulas given in [21]. More precisely, our formulas (48), (51) coincide with those given in [21], while (49), (50) have a different form in [21], namely:

$$\psi_{n,m}(t) \sim -\frac{(-1)^n}{2\pi}\left(\ln\frac{4\omega_2 t}{|n|} + \gamma - 2\right), \quad |m| = |n| \neq 0, \tag{52}$$

$$\psi_{n,m}(t) \sim -\frac{(-1)^n}{2\pi}\left(\ln\frac{16\omega_2 t}{|m^2 - n^2|} + \gamma - 4\right), \quad |m| \neq |n|. \tag{53}$$

The absolute value of the difference of the asymptotic expressions for $\psi_{n,n}$ from formulas (49), (52) is equal to

$$|\Delta\psi_{n,n}| \sim \frac{1}{2\pi}|\ln 4 + \gamma - 2| \approx \frac{0.0365}{2\pi}.$$

Thus, in the case $|m| = |n| \neq 0$ the absolute error between asymptotic solutions (49), (52) is constant, while the relative error tends to zero as $t \to \infty$. A similar situation occurs in the case $|m| \neq |n|$, namely, the absolute value of the difference of the asymptotic expressions for $\psi_{n,m}$ from formulas (50), (53) is constant and is equal to

$$|\Delta\psi_{n,m}| \sim \frac{1}{\pi}|\ln 4 + \gamma - 2| \approx \frac{0.073}{2\pi},$$

i.e., it is twice as large as in the case $|m| = |n| \neq 0$. Thus, we see that formulas (49), (50) are more accurate than formulas (52), (53).

As can be seen from (48)–(51), under a concentrated sinusoidal load with frequency $\omega_* = 2\omega_2$ the resonant wave propagates mainly along the diagonals of the lattice.

Fig. 9 shows the results of finite-difference calculations of displacements versus coordinates $n$, $m$ of the lattice under a concentrated sinusoidal load with resonant frequency $\omega_*$ ($t = 200$, $\tau = 0.07$). The calculations were performed using approximation (32) for the second time derivative. As can be seen in Fig. 9a, under the action of the load with frequency $\omega_* = 2\omega_2$, short-wave perturbations propagate mainly in diagonal directions. These are the so-called "star waves" [21]. As can be seen in Fig. 9b, under the action of the load with frequency $\omega_* = 2\sqrt{2}\omega_2$, resonant waves do not have a preferred direction of propagation. In Fig. 9a and b one can see the front of low-frequency wave propagating in all directions with the same velocity $c_2$ determined by (16).

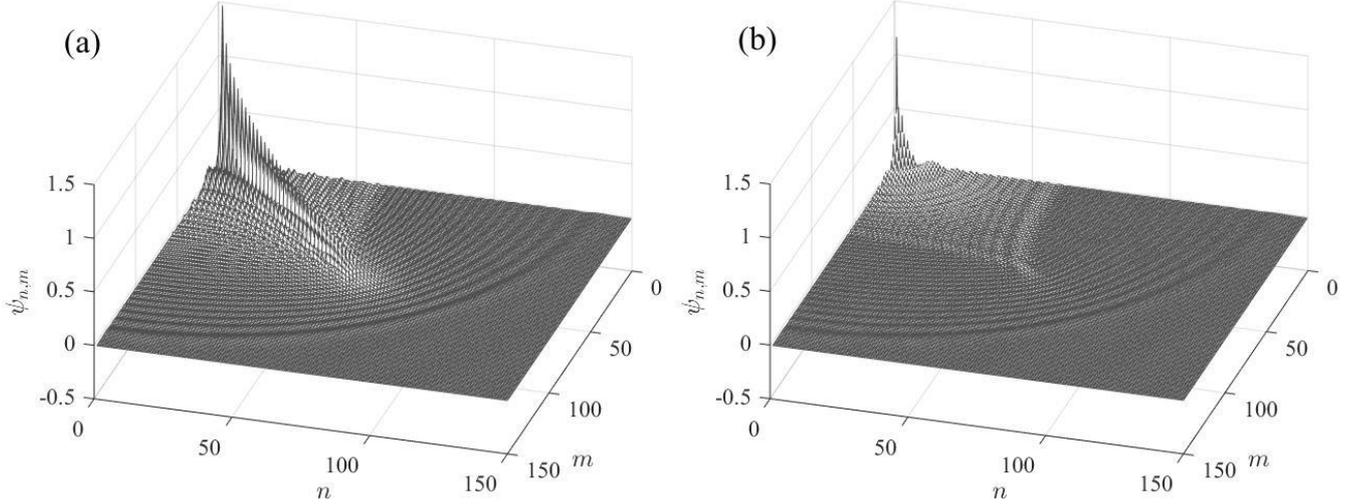

**Fig. 9**. Plots of displacements $\psi_{n,m}$ versus coordinates $n$, $m$ at the moment of time $t = 200$: (a) $\omega_* = 2\omega_2$, (b) $\omega_* = 2\sqrt{2}\omega_2$.

### 9. The plane problem. Load of the "center of dilatation" type

Let us continue the study of the plane problem described by Eqs. (1), (2). We use methods presented in Section 3 and Appendix A for solving the antiplane problem, but now we use them in a more complex situation of the plane problem. Namely, let us investigate the propagation of waves in a 2D lattice under a transient local load parallel to the plane of the lattice, see Fig. 10. In the lattice, we fix a square with the following vertices: $(n_0, m_0)$, $(n_0, m_0+1)$, $(n_0+1, m_0)$, $(n_0+1, m_0+1)$. Assume that the lattice is acted upon by external forces applied at the vertices of this square, and these forces have equal magnitudes and are directed from the center of symmetry of the square, see Fig. 10.

We say that such forces create a local load of the "center of dilatation" type. Let the dependence of the load on time be described by the Heaviside step function. In accordance with the above assumptions, we obtain the following formulas for projections $P_{n,m}$, $Q_{n,m}$ of the vectors of external forces appeared in (1), (2):

$$P_{n,m}(t) = \begin{cases} H(t), & \text{if } (n = n_0+1 \ \& \ m = m_0) \text{ or } (n = n_0+1 \ \& \ m = m_0+1), \\ -H(t), & \text{if } (n = n_0 \ \& \ m = m_0) \text{ or } (n = n_0 \ \& \ m = m_0+1), \\ 0, & \text{otherwise;} \end{cases}$$

$$Q_{n,m}(t) = \begin{cases} H(t), & \text{if } (n = n_0 \ \& \ m = m_0 + 1) \text{ or } (n = n_0 + 1 \ \& \ m = m_0 + 1), \\ -H(t), & \text{if } (n = n_0 \ \& \ m = m_0) \text{ or } (n = n_0 + 1 \ \& \ m = m_0), \\ 0, & \text{otherwise.} \end{cases}$$

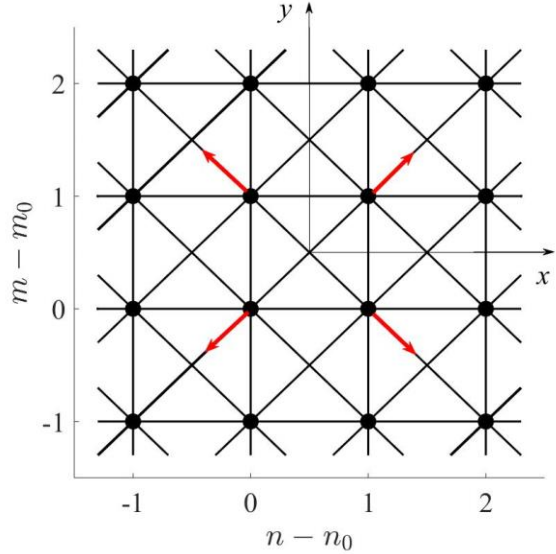

**Fig. 10.** Load of the "center of dilatation" type. The arrows show the direction and location of the external forces.

We apply to (1), (2) the Laplace transform with respect to time and the discrete Fourier transforms with respect to coordinates $n$, $m$. The Laplace–Fourier transforms of $u$ and $v$ are as follows:

$$u^{LF_nF_m}(p,q_x,q_y) = \frac{(p^2 + a_{22})P^{LF_nF_m} - a_{12}Q^{LF_nF_m}}{D(p,q_x,q_y)}, \quad v^{LF_nF_m}(p,q_x,q_y) = \frac{(p^2 + a_{11})Q^{LF_nF_m} - a_{21}P^{LF_nF_m}}{D(p,q_x,q_y)}, \quad (54)$$

where

$$P^{LF_nF_m}(p,q_x,q_y) = \frac{(e^{iq_x} - 1)(e^{iq_y} + 1)}{p}, \quad Q^{LF_nF_m}(p,q_x,q_y) = \frac{(e^{iq_x} + 1)(e^{iq_y} - 1)}{p}, \quad (55)$$

$$a_{11} = 3 - 2\cos q_x - \cos q_x \cos q_y, \quad a_{22} = 3 - 2\cos q_y - \cos q_x \cos q_y, \quad a_{12} = a_{21} = \sin q_x \sin q_y,$$
$$D(p,q_x,q_y) = p^4 + p^2(a_{11} + a_{22}) + a_{11}a_{22} - a_{21}a_{12} = D_1(p,q_x,q_y)D_2(p,q_x,q_y). \quad (56)$$

Here $D_1$ is defined by (19), $D_2$ is defined by (35).

Substituting (55), (56) into (54), we obtain:

$$u^{LF_nF_m}(p,q_x,q_y) = \frac{(e^{iq_x} - 1)(e^{iq_y} + 1)}{pD_1(p,q_x,q_y)}, \quad v^{LF_nF_m}(p,q_x,q_y) = \frac{(e^{iq_x} + 1)(e^{iq_y} - 1)}{pD_1(p,q_x,q_y)}. \quad (57)$$

Note that the denominator in formulas (57) contains the dispersion operator $D_1$ of the antiplane problem for longitudinal waves defined in Section 3. The functions $u$ and $v$, corresponding to $u^{LF_nF_m}$ and $v^{LF_nF_m}$ from (57), are calculated in Appendix B.

Let us define radial coordinate $r$ and radial displacement $u_r$ of the mass of the lattice with coordinates $n$, $m$ according to the formulas

$$r = \sqrt{(n-n_0-1/2)^2 + (m-m_0-1/2)^2}$$

and

$$u_r(n,m,t) = \frac{1}{r}[u_{n,m}(n-n_0-1/2) + v_{n,m}(m-m_0-1/2)].$$

Now we rewrite (B9) – (B11) in terms of the radial coordinate and radial displacement:

$$u_r(n,m,t) \sim \frac{1}{\omega_1^2}\left[\frac{\text{Ai}(\kappa_{r-1,1})\text{Ai}(\kappa_{r,1})}{(\omega_1 t/2)^{2/3}} + \frac{\{1-e^{-2[\omega_1 t/(r-3/2)-1]}\}}{\pi(r-1/2)} H[\omega_1 t - (r-3/2)]\right], \tag{58}$$

$$\dot{u}_r(n,m,t) \sim \frac{\text{Ai}^2(\kappa_{r-1,1}) - \text{Ai}^2(\kappa_{r,1})}{\omega_1(\omega_1 t/2)^{2/3}}, \tag{59}$$

$$\ddot{u}_r(n,m,t) \sim \frac{4}{\omega_1 t}\left[\text{Ai}(\kappa_{r,1})\text{Ai}'(\kappa_{r,1}) - \text{Ai}(\kappa_{r-1,1})\text{Ai}'(\kappa_{r-1,1})\right] \tag{60}$$

as $t \to \infty$ (or, what is the same, $r \to \infty$). Here $\kappa_{r,1}$ is given by (24).

In the case of the load of the "center of dilatation" type, it follows from (B6) that the value of the static radial displacement is determined by the formula:

$$u_r(n,m,t) \sim \frac{1}{\pi\omega_1^2(r-1/2)}. \tag{61}$$

## 10. The plane problem. Load of the "center of rotation" type

Let us investigate the propagation of waves in a 2D lattice under a transient local load parallel to the plane of the lattice, see Fig. 11. As in Section 9, in the lattice, we fix a square with the following vertices: $(n_0, m_0)$, $(n_0, m_0+1)$, $(n_0+1, m_0)$, $(n_0+1, m_0+1)$. Assume that the lattice is acted upon by external forces applied at the vertices of this square, and these forces have equal magnitudes and are directed as shown in Fig. 11. We say that such forces create a local load of the "center of rotation" type. Let the dependence of the load on time be described by the Heaviside step function.

In accordance with the above assumptions, we obtain the following formulas for projections $P_{n,m}$, $Q_{n,m}$ of the vectors of external forces appeared in (1), (2):

$$P_{n,m}(t) = \begin{cases} H(t), & \text{if } (n=n_0 \,\&\, m=m_0) \text{ or } (n=n_0+1 \,\&\, m=m_0), \\ -H(t), & \text{if } (n=n_0 \,\&\, m=m_0+1) \text{ or } (n=n_0+1 \,\&\, m=m_0+1), \\ 0, & \text{otherwise}; \end{cases}$$

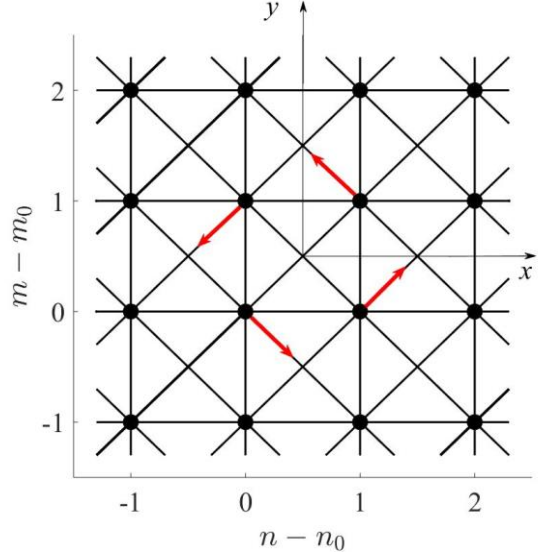

**Fig. 11.** Load of the "center of rotation" type. The arrows show the direction and location of the external forces.

$$Q_{n,m}(t) = \begin{cases} H(t), & \text{if } (n = n_0 + 1 \ \& \ m = m_0) \text{ or } (n = n_0 + 1 \ \& \ m = m_0 + 1), \\ -H(t), & \text{if } (n = n_0 \ \& \ m = m_0) \text{ or } (n = n_0 \ \& \ m = m_0 + 1), \\ 0, & \text{otherwise.} \end{cases}$$

In the case of the load of the "center of rotation" type, the Laplace–Fourier transforms of $P_{n,m}$ and $Q_{n,m}$ are as follows:

$$P^{LF_nF_m}(p, q_x, q_y) = \frac{(e^{iq_x} + 1)(1 - e^{iq_y})}{p}, \quad Q^{LF_nF_m}(p, q_x, q_y) = \frac{(e^{iq_x} - 1)(1 + e^{iq_y})}{p}. \tag{62}$$

Substituting (56), (62) into (54), we obtain:

$$u^{LF_nF_m}(p, q_x, q_y) = \frac{(e^{iq_x} + 1)(1 - e^{iq_y})}{pD_2(p, q_x, q_y)}, \quad v^{LF_nF_m}(p, q_x, q_y) = \frac{(e^{iq_x} - 1)(1 + e^{iq_y})}{pD_2(p, q_x, q_y)}, \tag{63}$$

where $D_2$ is defined by (35).

From (63) we can obtain asymptotic formulas for displacements $u_{n,1}$, $v_{n,1}$ and their time derivatives as $t \to \infty$ (or $n \to \infty$). For this, one can transform (63) in a way, similar to those used in Appendix B. Omitting the calculations, we have:

$$u_{n,1}(t) \sim 0, \quad \dot{u}_{n,1}(t) \sim 0, \quad \ddot{u}_{n,1}(t) \sim 0,$$

$$v_{n,1}(t) \sim \frac{1}{\omega_2^2}\left[ J_{n-1}(\omega_2 t) J_n(\omega_2 t) + \frac{\{1 - e^{-2[\omega_2 t/(n-3/2)-1]}\}}{\pi(n - 1/2)} H[\omega_2 t - (n - 3/2)] \right], \tag{64}$$

$$\dot{v}_{n,1}(t) \sim \frac{1}{\omega_2}[J_{n-1}^2(\omega_2 t) - J_n^2(\omega_2 t)], \tag{65}$$

$$\ddot v_{n,1}(t) \sim 2[J_{n-1}(\omega_2 t)J'_{n-1}(\omega_2 t) - J_n(\omega_2 t)J'_n(\omega_2 t)], \qquad (66)$$

where $\omega_2$ is defined by (16).

Using (A4) and (A7), we obtain from (64)–(66) the following solution in terms of the Airy function:

$$v_{n,1}(t) \sim \frac{1}{\omega_2^2}\left[\frac{\mathrm{Ai}(\kappa_{n-1,2})\mathrm{Ai}(\kappa_{n,2})}{(\omega_2 t/2)^{2/3}} + \frac{\{1-e^{-2[\omega_2 t/(n-3/2)-1]}\}}{\pi(n-1/2)}H[\omega_2 t-(n-3/2)]\right], \qquad (67)$$

$$\dot v_{n,1}(t) \sim \frac{\mathrm{Ai}^2(\kappa_{n-1,2}) - \mathrm{Ai}^2(\kappa_{n,2})}{\omega_2(\omega_2 t/2)^{2/3}}, \qquad (68)$$

$$\ddot v_{n,1}(t) \sim \frac{4}{\omega_2 t}\left[\mathrm{Ai}(\kappa_{n,2})\mathrm{Ai}'(\kappa_{n,2}) - \mathrm{Ai}(\kappa_{n-1,2})\mathrm{Ai}'(\kappa_{n-1,2})\right], \qquad (69)$$

where $\kappa_{n,2}$ is defined by (39).

Let us define tangential displacement $u_\theta$ of the mass of the lattice with coordinates $n$, $m$ according to the formula

$$u_\theta(n,m,t) = \frac{1}{r}[v_{n,m}(n-n_0-1/2) - u_{n,m}(m-m_0-1/2)].$$

Now we rewrite (67)–(69) in terms of the radial coordinate $r$ and tangential displacement:

$$u_\theta(n,m,t) = \frac{1}{\omega_2^2}\left[\frac{\mathrm{Ai}(\kappa_{r-1,2})\mathrm{Ai}(\kappa_{r,2})}{(\omega_2 t/2)^{2/3}} + \frac{\{1-e^{-2[\omega_2 t/(r-3/2)-1]}\}}{\pi(r-1/2)}H[\omega_2 t-(r-3/2)]\right], \qquad (70)$$

$$\dot u_\theta(n,m,t) = \frac{\mathrm{Ai}^2(\kappa_{r-1,2}) - \mathrm{Ai}^2(\kappa_{r,2})}{\omega_2(\omega_2 t/2)^{2/3}}, \qquad (71)$$

$$\ddot u_\theta(n,m,t) = \frac{4}{\omega_2 t}\left[\mathrm{Ai}(\kappa_{r,2})\mathrm{Ai}'(\kappa_{r,2}) - \mathrm{Ai}(\kappa_{r-1,2})\mathrm{Ai}'(\kappa_{r-1,2})\right]. \qquad (72)$$

In the case of the load of the "center of rotation" type, the value of the static tangential displacement is determined by the formula:

$$u_\theta(n,m,t) = \frac{1}{\pi\omega_2^2(r-1/2)}. \qquad (73)$$

## 11. Comparison of finite-difference and analytical solutions for the plane problem

Eqs. (1), (2) are solved by the finite difference method according to an explicit scheme using approximation (32) for time derivative. Numerical calculations, the results of which are shown in Figs. 12–19, were carried out for the following values of parameters: $n_0 = m_0 = 75$, and $\tau = 0.07$.

Figs. 12, 14 – 16 show the results of numerical calculations for radial displacement $u_r(n,m,t)$ and its velocity for a step load of the "center of dilatation" type. Figs. 13, 17 – 19 show the results of numerical calculations for tangential displacement $u_\theta(n,m,t)$ and its velocity for a step load of the "center of rotation" type.

Figs. 12, 13 show the results of numerical calculations of the radial and tangential displacements and their velocities at the points of the lattice with coordinates $0 \leq m \leq 150$, $0 \leq n \leq 150$ at the moment of time $t = 50$. As can be seen in Figs. 12, 13, under the load of the "center of dilatation" type, a longitudinal low-frequency wave is predominantly formed in the lattice, propagating with velocity $c_1$, while under the load of the "center of rotation" type, a shear low-frequency wave is predominantly formed, propagating with velocity $c_2$.

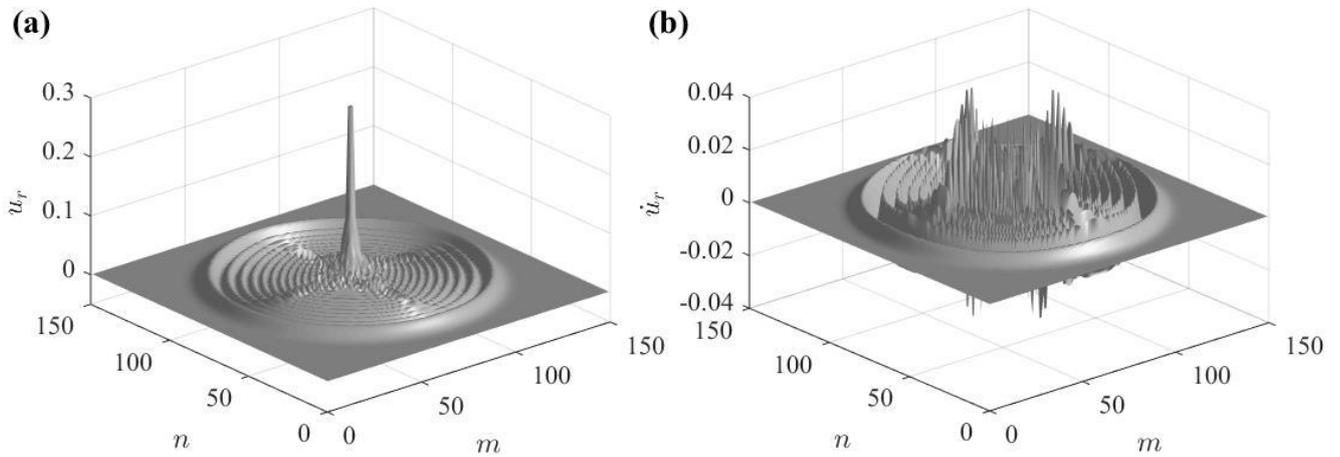

**Fig. 12.** Plots of the radial displacement (a) and its velocity (b) versus coordinates at the moment of time $t = 50$ under a step load of the "center of dilatation" type.

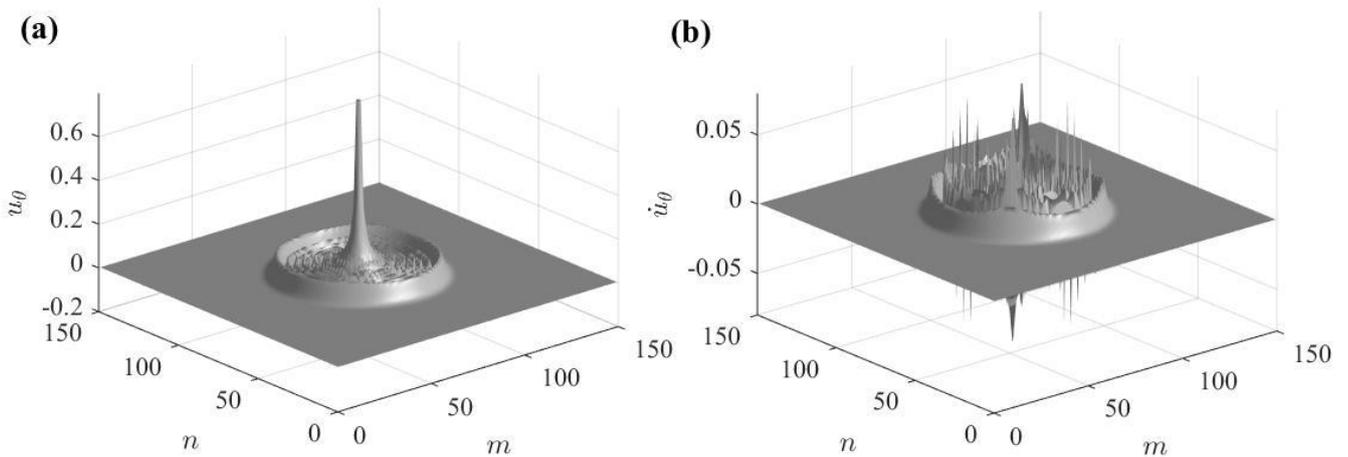

**Fig. 13.** Plots of the tangential displacement (a) and its velocity (b) versus coordinates at the moment of time $t = 50$ under a step load of the "center of rotation" type.

Figs. 14 – 19 show the plots of radial and tangential displacements, their velocities, and accelerations versus time at the points with coordinates $n=n_0+31$, $m=m_0+1$ on the axis of the loading, and coordinates $n=n_0+22$, $m=m_0+22$ on the diagonal. These points are located approximately at the same distance from the place of loading. The thin curves correspond to the finite-difference solutions; the thick solid curves correspond to analytical solutions. In Figs. 14 – 16, the vertical dashed lines correspond to the arrival time of longitudinal wave propagating with velocity $c_1$. In Figs. 17 – 19, the vertical dashed lines correspond to the arrival time of shear wave propagating with velocity $c_2$.

From (61), (73) and Figs. 14, 17, it can be seen that radial and tangential displacements behind the front of the low-frequency waves oscillate with respect to their static values, which decrease proportionally to the distance from the place of loading.

As can be seen in Figs. 12, 14 – 16, in the case of the load of the "center of dilatation" type, the amplitudes of high-frequency oscillations of the radial velocities and accelerations of masses, formed behind the front of the low-frequency wave, on the axis of loading are significantly greater than on the diagonal. On the contrary, it can be seen in Figs. 13, 17 – 19 that, in the case of the load of the "center of rotation" type, the amplitudes of high-frequency oscillations of the tangential velocities and accelerations of masses, formed behind the front of the low-frequency wave, on the axis of loading are significantly less than on the diagonal.

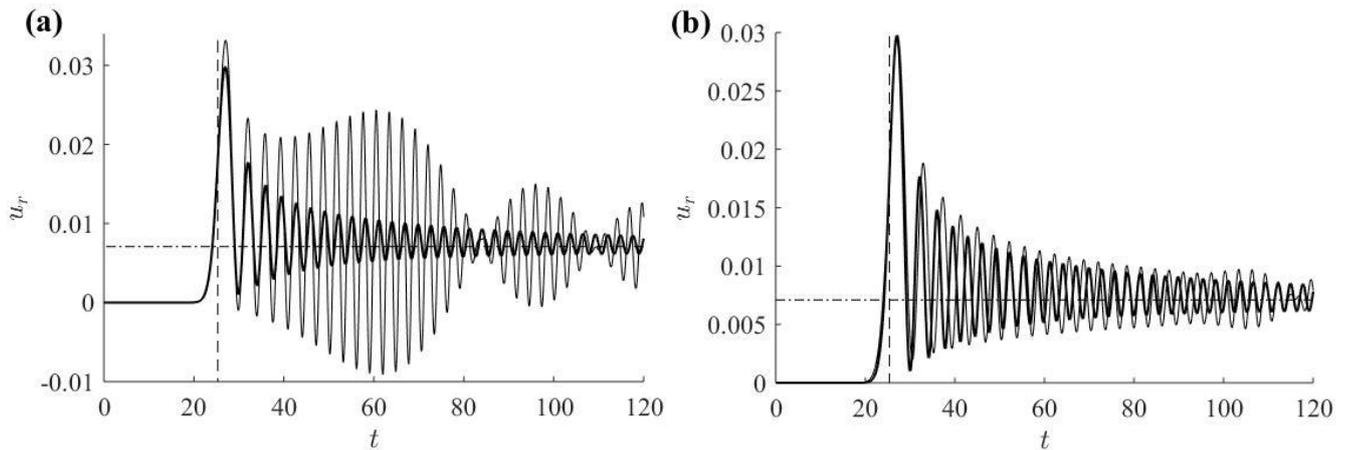

**Fig. 14.** Plots of the radial displacement versus time under a step load of the "center of dilatation" type: (a) $n=n_0+31$, $m=m_0+1$; (b) $n=n_0+22$, $m=m_0+22$. Thin curves correspond to the finite-difference solutions, thick curves correspond to analytical solution (58), dash-dotted lines correspond to the static values of displacement calculated by (61).

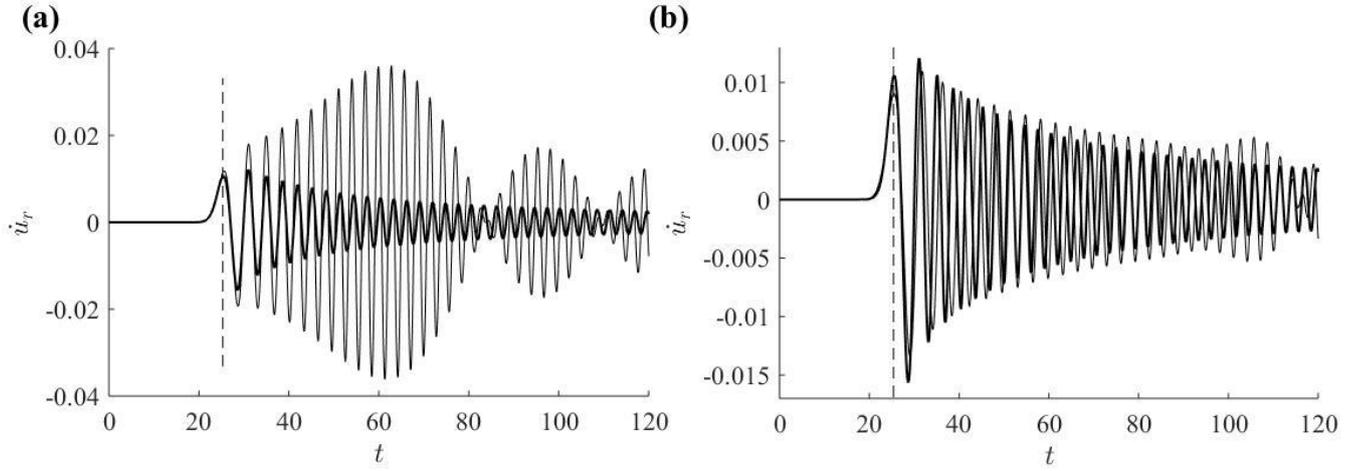

**Fig. 15.** Plots of the radial velocities of masses versus time under a step load of the "center of dilatation" type: (a) $n = n_0 + 31$, $m = m_0 + 1$; (b) $n = n_0 + 22$, $m = m_0 + 22$. Thin curves correspond to the finite-difference solutions, thick curves correspond to analytical solution (59).

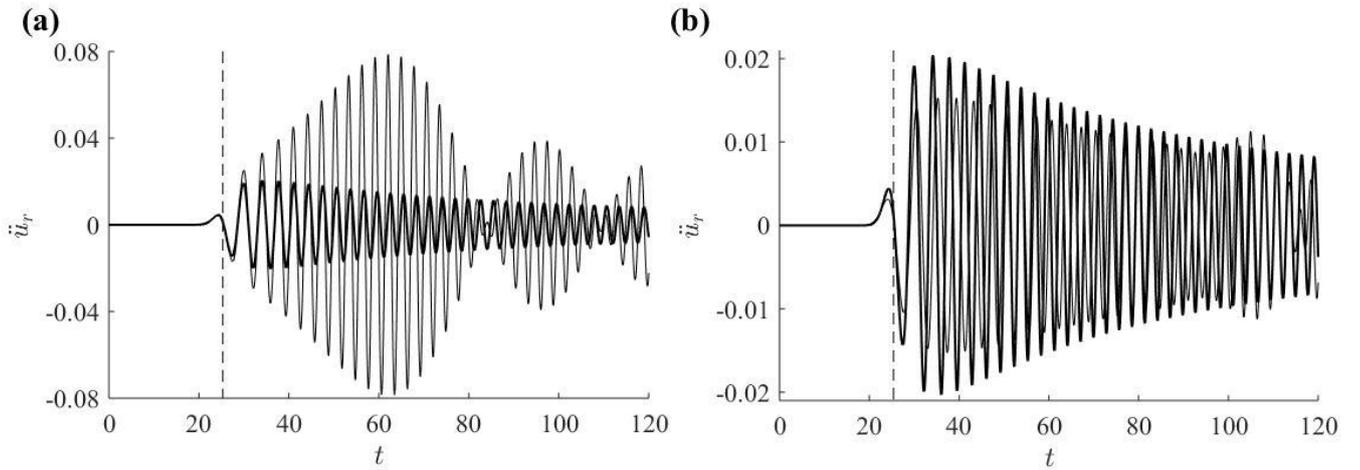

**Fig. 16.** Plots of the radial accelerations of masses versus time under a step load of the "center of dilatation" type: (a) $n = n_0 + 31$, $m = m_0 + 1$; (b) $n = n_0 + 22$, $m = m_0 + 22$. Thin curves correspond to the finite-difference solutions, thick curves correspond to analytical solution (60).

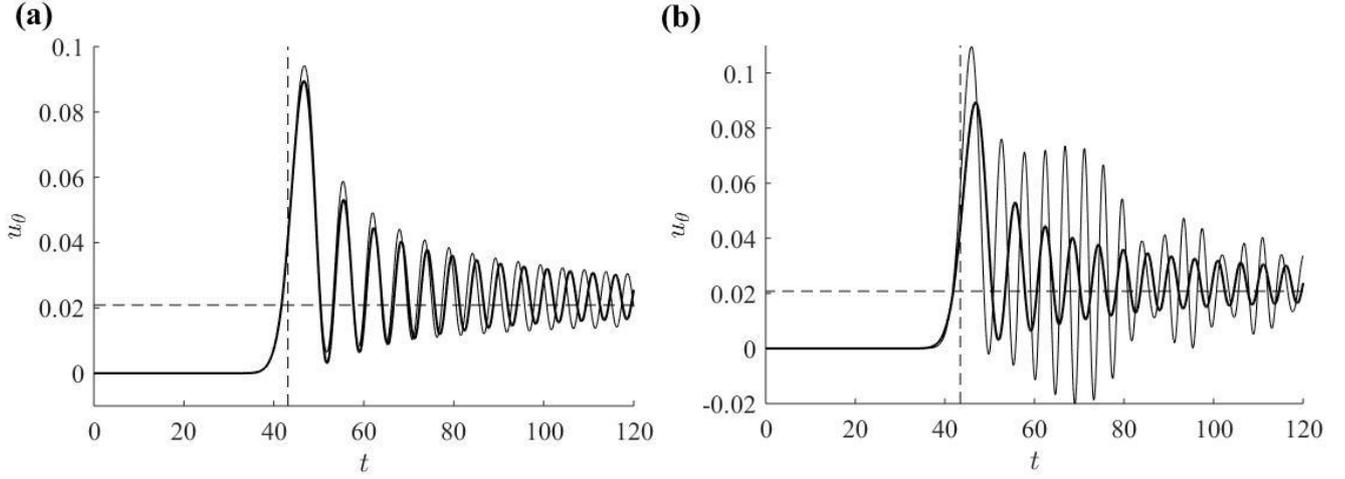

**Fig. 17.** Plots of the tangential displacement versus time under a step load of the "center of rotation" type: (a) $n = n_0 + 31$, $m = m_0 + 1$; (b) $n = n_0 + 22$, $m = m_0 + 22$. Thin curves correspond to the finite-difference solutions, thick curves correspond to analytical solution (70), dash-dotted lines correspond to the static values of displacement calculated by (73).

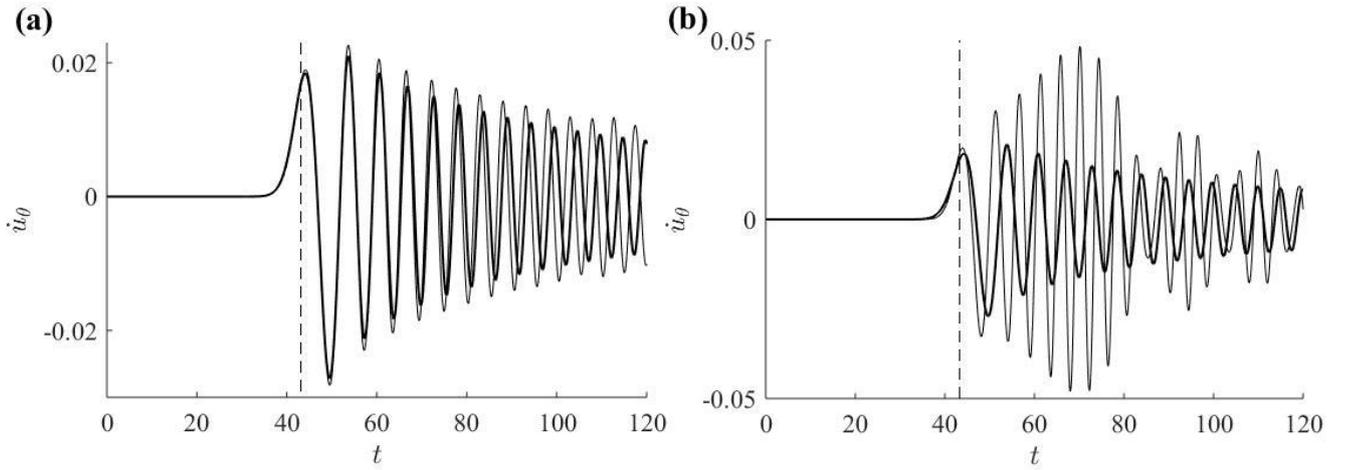

**Fig. 18.** Plots of the tangential velocities of masses versus time under a step load of the "center of rotation" type: (a) $n = n_0 + 31$, $m = m_0 + 1$; (b) $n = n_0 + 22$, $m = m_0 + 22$. Thin curves correspond to the finite-difference solutions, thick curves correspond to analytical solution (71).

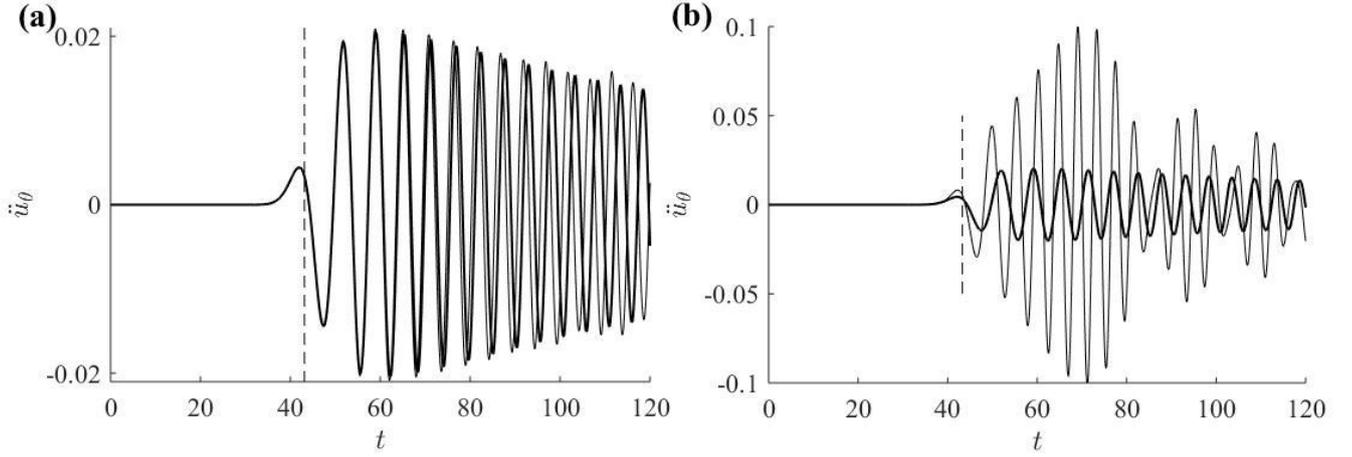

**Fig. 19.** Plots of the tangential accelerations of masses versus time under a step load of the "center of rotation" type: (a) $n = n_0 + 31$, $m = m_0 + 1$; (b) $n = n_0 + 22$, $m = m_0 + 22$. Thin curves correspond to the finite-difference solutions, thick curves correspond to analytical solution (72).

**12. Conclusion**

In this article, it is shown that equations describing 2D plane motion of a square lattice, in which point masses are connected by elastic springs in axial and diagonal directions, can be represented in the form of two linearly independent wave equations, each of which contains one unknown function only. One of those equations describes the propagation of shear waves in the lattice, in which the masses are connected by springs in axial directions only. The other equation describes the propagation of longitudinal waves in the lattice, in which the masses are connected by springs in both axial and diagonal directions.

Asymptotic solutions are obtained that describe the propagation of antiplane perturbations in 2D square lattices of two types (where the masses are connected only in the axial directions, and where the masses are connected in both the axial and diagonal directions) under a concentrated step load in the direction orthogonal to the plane of the lattice. It is shown that, under such load, the amplitudes of the displacements are proportional to $\ln(t/r)$ as $t \to \infty$, $r \neq 0$.

Asymptotic solutions are obtained that describe the propagation of perturbations in 2D square lattices in the plane formulation under step loads of the "center of dilatation" and "center of rotation" types. Under such loads, displacements oscillate with respect to their static values, which decrease as $1/r$ when $r \to \infty$. The amplitude of these oscillations decreases with increasing time (distance) as $t^{-2/3}$ ($r^{-2/3}$). It is shown that under the load of the "center of dilatation" type, mainly longitudinal waves are formed in the lattice, and the maximum perturbations are achieved in the radial direction. Similarly, it is shown that under the load of the "center of rotation" type, mainly shear waves are formed in the lattice, and the maximum perturbations are achieved in the tangential direction.

It is shown that, in the plane and antiplane formulations of the 2D problems under consideration, the following effects are observed in the lattice as $t \to \infty$ ($r \to \infty$): the maximum amplitude of the velocities of masses decrease as $t^{-2/3}$ ($r^{-2/3}$); the quasi-front zone, i.e., zone, where perturbations vary from zero to maximum, expands as $t^{1/3}$ ($r^{1/3}$).

The method proposed in this article is useful for the analytical study of the propagation of transient waves in discrete-periodic media.

Analytical and numerical solutions for 2D lattices are compared with each other. It is shown that they agree with each other qualitatively and quantitatively for both the plane and antiplane problems. This agreement occurs at a finite exposure time or, which is the same, at a finite distance from the place of loading.

In this article, solutions are obtained for a step load. In many seismic problems, the study of perturbations under an impulse load is even more important. Solutions for perturbations under an impulse load can be obtained from the solutions presented in this article using Duhamel's integral.

**Acknowledgement**

**Appendix A. Calculation of the inverse Laplace and Fourier transforms of function $\varphi^{LF_nF_m}$ given by formula (18)**

Applying the inverse discrete Fourier transform with respect to $q_y$ to $\varphi^{LF_nF_m}$, we obtain:

$$\varphi_m^{LF_n}(p, q_x) = \frac{1}{\pi p B_1} \int_0^\pi \frac{\cos(q_y |m|) dq_y}{1 - (2\cos q_x + 1) B_1^{-1} \cos q_y}.$$

Using [28], for $m = 0$ we obtain:

$$\varphi_0^{LF_n}(p, q_x) = \frac{1}{p\sqrt{B_1^2 - (2\cos q_x + 1)^2}}.$$

Applying the inverse discrete Fourier transform with respect to $q_x$ to $\varphi_0^{LF_n}$ and using the formula for the Laplace transform of the derivative, we obtain:

$$\dot{\varphi}_0^L(p) = \frac{2}{\pi} \int_0^{\pi/2} \frac{\cos(2qn)dq}{\sqrt{(p^2 + 6\sin^2 q)(p^2 - 2\sin^2 q + 6)}}. \tag{A1}$$

Let us derive the asymptotics of perturbations for large values of time $t$. As is known, the condition $t \to \infty$ in the time domain corresponds to the condition $p \to 0$ in the $p$ domain. Replacing the integral in (A1) by an asymptotically equivalent integral for $p \to 0$ and large values of $n$, we obtain:

$$\dot{\varphi}_{0,n}^L(p) \sim \frac{1}{\pi\omega_1} \int_0^{\pi/2} \frac{\cos(2qn)dq}{\sqrt{p^2 + 4\omega_1^2 \sin^2 q}}.$$

Here $f \sim g$ means that $f$ is asymptotically equal to $g$. Using [27] and [28], we obtain the following asymptotic formula:

$$\dot{\varphi}_{n,0}(t) \sim \frac{1}{\pi\omega_1} \int_0^{\pi/2} J_0(2\omega_1 t \sin q)\cos(2qn)dq = \frac{1}{2\omega_1} J_n^2(\omega_1 t) \quad \text{as} \quad t \to \infty \text{ and } n \to \infty. \tag{A2}$$

Here $J_n$ is the Bessel function of the first kind of integer order $n$ [28].

Differentiating (A2) with respect to time, we find the asymptotic behavior of acceleration $\ddot{\varphi}_{n,0}$:

$$\ddot{\varphi}_{n,0} \sim J_n(\omega_1 t) J_n'(\omega_1 t) \quad \text{as} \quad t \to \infty \text{ and } n \to \infty. \tag{A3}$$

Here the prime denotes the derivative with respect to the argument of the function.

Next, we need the following asymptotic formula for the Bessel function:

$$J_n(\omega t) \sim \frac{\text{Ai}(\kappa)}{(\omega t/2)^{1/3}} \quad \text{as} \quad t \to \infty \text{ and } n \to \infty. \tag{A4}$$

Here

$$\text{Ai}(\kappa) = \frac{1}{\pi} \int_0^\infty \cos\left(\kappa y + \frac{y^3}{3}\right) dy$$

is the Airy function and

$$\kappa = \frac{n - \omega t}{(\omega t/2)^{1/3}}.$$

Formula (A4) was first obtained in [29] and is known as a Nicholson-type formula (see page 142 in [30] or pages 190 and 249 in [31]).

Using (A4), rewrite (A2) in the following form:

$$\dot{\varphi}_{n,0}(t) \sim \frac{1}{2\omega_1} \left[\frac{\text{Ai}(\kappa_{n,1})}{(\omega_1 t/2)^{1/3}}\right]^2 \quad \text{as} \quad t \to \infty \text{ and } n \to \infty, \tag{A5}$$

where

$$\kappa_{n,1} = \frac{n - \omega_1 t}{(\omega_1 t/2)^{1/3}}.$$

Now we use the following asymptotic formula valid in the vicinity of point $n = \omega t$:

$$J'_n(\omega t) \sim -\frac{\text{Ai}'(\kappa)}{(\omega t/2)^{2/3}} \quad \text{as} \quad t \to \infty \quad \text{and} \quad n \to \infty. \tag{A6}$$

Formula (A6) was derived in [32]. Using (A4) and (A6), from (A3) we obtain the following asymptotic formula for $\ddot{\varphi}_{n,0}$ in terms of the Airy function:

$$\ddot{\varphi}_{n,0}(t) \sim -\frac{2\text{Ai}(\kappa_{n,1})\text{Ai}'(\kappa_{n,1})}{\omega_1 t} \quad \text{as} \quad t \to \infty. \tag{A7}$$

Solutions (A5), (A7) obtained in terms of the Airy function are less accurate than solutions (A2), (A3) obtained in terms of the Bessel functions. However the advantage of solutions (A5), (A7), over (A2), (A3) is that they explicitly describe the degree of attenuation of long-wave perturbations with increasing time or distance in the vicinity of the quasi-front $n = c_1 t$ of the longitudinal wave.

Now let's calculate displacement $\varphi_{n,0}$. Integrating (A2) with respect to time and using [28], we obtain the following asymptotic formula for $\varphi_0^{F_n}$:

$$\varphi_{n,0}(t) \sim \frac{1}{4\pi\omega_1^2} \sum_{k=0}^{\infty} \int_0^{\pi/2} \frac{J_{2k+1}(2\omega_1 t \sin q)\cos(2qn)}{\sin q} dq \quad \text{as} \quad t \to \infty.$$

Since the main contribution to the integral is made by an arbitrarily small neighborhood of point $q = 0$, then we replace the integral over $(0, \pi/2)$ by the integral over $(0, \varepsilon)$, where $\varepsilon$ is a small positive number, and we replace $\sin q$ by $q$ in the integrand. Then, for the same reason, we replace the integral over $(0, \varepsilon)$ by the integral over $(0, \infty)$. Using [28], we obtain:

$$\varphi_{n,0}(t) \sim \frac{1}{4\pi\omega_1^2} \sum_{k=0}^{\infty} \int_0^{\infty} \frac{J_{2k+1}(2\omega_1 tq)\cos(2qn)}{q} dq = \frac{H(\omega_1 t - n)}{2\pi\omega_1^2} \ln \text{ctg}\left[\frac{1}{2}\arcsin\left(\frac{n}{\omega_1 t}\right)\right]$$
$$= \frac{H(\omega_1 t - n)}{2\pi\omega_1^2} \ln\left[\frac{\omega_1 t}{n} + \sqrt{\left(\frac{\omega_1 t}{n}\right)^2 - 1}\right]. \tag{A8}$$

Differentiating both sides of this formula in time, yields:

$$\dot{\varphi}_{n,0}(t) \sim \frac{H(\omega_1 t - n)}{2\pi\omega_1 \sqrt{\omega_1^2 t^2 - n^2}}. \tag{A9}$$

The above asymptotic expressions (A8), (A9) for $\varphi_{n,0}$, $\dot{\varphi}_{n,0}$ are valid for $n \neq 0$. Let us find similar asymptotic expression $\varphi_{n,0}$ for $n = 0$. Letting $p \to 0$ and $n = 0$, we obtain from (A1):

$$\varphi_{0,0}^{L}(p) \sim \frac{1}{3\pi p} \int_{0}^{\pi/2} \frac{dq}{\sqrt{[p^2/(4\omega_1^2)+\sin^2 q][1-\sin^2 q/3]}}.$$

Let us split this integral into two integrals:

$$\varphi_{0,0}^{L}(p) \sim \frac{1}{3\pi p}(I_1+I_2) = \frac{1}{3\pi p}\left(\int_{0}^{\varepsilon} + \int_{\varepsilon}^{\pi/2}\right),$$

where $\varepsilon$ is a small positive number. We calculate the first integral:

$$I_1 \sim \int_{0}^{\varepsilon} \frac{dq}{\sqrt{p^2/(4\omega_1^2)+q^2}} = \ln\left(\sqrt{1+\frac{4\varepsilon^2\omega_1^2}{p^2}}+\frac{2\varepsilon\omega_1}{p}\right) \sim \ln\frac{4\varepsilon\omega_1}{p}.$$

Then we calculate the second integral:

$$I_2 \sim \int_{\varepsilon}^{\pi/2} \frac{dq}{\sin q\sqrt{1-\sin^2 q/3}} \sim \frac{1}{2}\ln\frac{\sqrt{1-\sin^2 q/3}+\cos\varepsilon}{\sqrt{1-\sin^2 q/3}-\cos\varepsilon} \sim \ln\frac{\sqrt{6}}{\varepsilon}.$$

As a result, we get the following formula:

$$\varphi_{0,0}^{L}(p) \sim \frac{1}{2\pi p\omega_1^2}\ln\frac{4\sqrt{6}\omega_1}{p} \quad \text{as} \quad p \to 0.$$

Using [27], we calculate the inverse Laplace transform from both sides of the last formula. As a result, we obtain the asymptotics of displacement $\varphi_{0,0}$ at the point of load application:

$$\varphi_{0,0}(t) \sim \frac{1}{2\pi\omega_1^2}\left[\ln\left(4\sqrt{6}\omega_1 t\right)+\gamma\right] \quad \text{as} \quad t \to \infty.$$

Here $\gamma \approx 0.577$ is the Euler constant.

**Appendix B. Calculation of the inverse Laplace and Fourier transforms of functions $u^{LF_n F_m}$, $v^{LF_n F_m}$ given by formulas (57)**

Apply the inverse Fourier transform with respect to variable $q_y$ to functions $u^{LF_n F_m}$, $v^{LF_n F_m}$ (57), and put $m=1$. This yields:

$$u_1^{LF_n}(p,q_x) = \frac{2(e^{iq_x}-1)}{p\sqrt{p^2+6\sin^2(q_x/2)}\left[\sqrt{p^2+6\sin^2(q_x/2)}+\sqrt{p^2+6-2\sin^2(q_x/2)}\right]},$$

$$v_1^{LF_n}(p,q_x) = \frac{2(e^{iq_x}+1)}{p\sqrt{p^2+6-2\sin^2(q_x/2)}\left[\sqrt{p^2+6\sin^2(q_x/2)}+\sqrt{p^2+6-2\sin^2(q_x/2)}\right]}.$$

We apply the inverse Fourier transform with respect to variable $q_x$ to the last two formulas. This yields:

$$u_{n,1}^L(p) = \frac{4}{\pi p} \int_0^{\pi/2} \frac{[\cos 2(n-1)q - \cos 2nq]}{\sqrt{p^2 + 6\sin^2 q}\left[\sqrt{p^2 + 6\sin^2 q} + \sqrt{p^2 + 6 - 2\sin^2 q}\right]} dq,$$

$$v_{n,1}^L(p) = \frac{4}{\pi p} \int_0^{\pi/2} \frac{[\cos 2(n-1)q + \cos 2nq]}{\sqrt{p^2 + 6 - 2\sin^2 q}\left[\sqrt{p^2 + 6\sin^2 q} + \sqrt{p^2 + 6 - 2\sin^2 q}\right]} dq.$$

Next, we proceed in the same way as we did in solving the antiplane problem (see Appendix A). For $p \to 0$ and large values of $n$, the following asymptotic formulas hold true.

$$u_{n,1}^L(p) \sim \frac{2}{\pi p \omega_1} \int_0^{\pi/2} \frac{[\cos 2(n-1)q - \cos 2nq]}{\sqrt{p^2 + 4\omega_1^2 \sin^2 q}} dq, \quad v_{n,1}^L(p) \sim \frac{2}{3\pi p} \int_0^{\pi/2} [\cos 2(n-1)q + \cos 2nq] dq = 0, \quad \text{(B1)}$$

where $\omega_1$ is defined by (16). As can be seen from (B1), the asymptotics of displacement $v_{n,1}^L$ is equal to zero.

Now let us investigate the asymptotics of displacement $u_{n,1}$ and its time derivatives. We calculate the inverse Laplace transform of velocity of the mass $\dot{u}_{n,1}^L$:

$$\dot{u}_{n,1}(t) \sim \frac{2}{\pi \omega_1} \int_0^{\pi/2} [\cos 2(n-1)q - \cos 2nq] J_0\left(2t\omega_1 \sin q\right) dq = \frac{1}{\omega_1}\left[J_{n-1}^2(\omega_1 t) - J_n^2(\omega_1 t)\right]. \quad \text{(B2)}$$

Differentiating (B2) with respect to time, we find the asymptotics of the acceleration of the mass:

$$\ddot{u}_{n,1}(t) \sim 2\left[J_{n-1}(\omega_1 t) J'_{n-1}(\omega_1 t) - J_n(\omega_1 t) J'_n(\omega_1 t)\right]. \quad \text{(B3)}$$

Let us now calculate the asymptotic behavior of displacement $u_{n,1}$. Using [28], we rewrite (B2) in another form:

$$\dot{u}_{n,1}(t) \sim \frac{2}{\pi \omega_1} \int_0^{\pi/2} [J_{2(n-1)}\left(2\omega_1 t \sin q\right) - J_{2n}\left(2\omega_1 t \sin q\right)] dq. \quad \text{(B4)}$$

Integrating (B4) over time, we obtain the asymptotics for displacement:

$$u_{n,1}(t) \sim \frac{2}{\pi \omega_1^2} \int_0^{\pi/2} \frac{J_{2n-1}\left(2\omega_1 t \sin q\right)}{\sin q} dq. \quad \text{(B5)}$$

Since the main contribution to the integral is made by an arbitrarily small neighborhood of point $q = 0$, then we replace the integral over $(0, \pi/2)$ by the integral over $(0, \varepsilon)$, where $\varepsilon$ is a small positive number, and we replace $\sin q$ by $q$ in the integrand. Then, for the same reason, we replace the integral over $(0, \varepsilon)$ by the integral over $(0, \infty)$. As a result, we obtain the contribution of the singular point $q = 0$ to the asymptotics of $u_{n,1}$:

$$u_{n,1}(t) \sim \frac{2}{\pi\omega_1^2} \int_0^\infty \frac{J_{2n-1}(2\omega_1 tq)}{q} dq = \frac{1}{\pi\omega_1^2(n-1/2)}. \tag{B6}$$

Now let us study the asymptotics of $u_{n,1}$ as a whole. For this, in (B5), we represent the Bessel function in the form of a series using [28] and integrate that series term by term. This yields

$$u_{n,1}(t) \sim \frac{1}{\omega_1^2}\left[J_{n-1}(\omega_1 t)J_n(\omega_1 t) + \sum_{k=0}^\infty \frac{(-1)^k(\omega_1 t/2)^{2k+2n-1}(2n+2k-2)!}{k!(2n+k-1)!(n+k-1)!(n+k)!}\right]. \tag{B7}$$

Put by definition

$$F(n,\omega_1 t) = \frac{1}{\omega_1^2}\sum_{k=0}^\infty \frac{(-1)^k(\omega_1 t/2)^{2k+2n-1}(2n+2k-2)!}{k!(2n+k-1)!(n+k-1)!(n+k)!}$$

and find an approximate formula for this function. From (B5) and (B7) we obtain:

$$F(n,\omega_1 t) = \frac{1}{\omega_1^2}\left[\frac{2}{\pi}\int_0^{\pi/2}\frac{J_{2n-1}(2\omega_1 t\sin q)}{\sin q}dq - J_{n-1}(\omega_1 t)J_n(\omega_1 t)\right].$$

The thin curve in Fig. B1a shows a plot of function $F(n,\omega_1 t)$ versus time $t$ for $n=11$. To do this, the integral is calculated numerically with precision $10^{-7}$ using MATLAB. The plot of the function $F(n,\omega_1 t)$ in Fig. B1a resembles the plot of the well-known function $[1-e^{-b(t-a)}]H(t-a)$. Choosing the parameters $a$, $b$ so that the last function is as close as possible to $F(n,\omega_1 t)$, we get the following function:

$$F_1(n,\omega_1 t) = \frac{\{1-e^{-2[\omega_1 t/(n-3/2)-1]}\}}{\omega_1^2 \pi(n-1/2)}H[\omega_1 t-(n-3/2)].$$

The thick curve in Fig. B1a shows a plot of function $F_1(n,\omega_1 t)$ versus time $t$ for $n=11$. As can be seen in Fig. B1a, the plots of functions $F(n,\omega_1 t)$ and $F_1(n,\omega_1 t)$ are close to each other both in the vicinity of the front of the longitudinal wave, i.e. for $t \approx (n-1/2)/\omega_1$, and for $t\to\infty$. From this we conclude that $F_1(n,\omega_1 t)$ can be considered as a reasonable approximation for $F(n,\omega_1 t)$. Replacing $F(n,\omega_1 t)$ by $F_1(n,\omega_1 t)$ in (B7), we obtain the following approximate formula for displacement:

$$u_{n,1}(t) \sim \frac{1}{\omega_1^2}\left[J_{n-1}(\omega_1 t)J_n(\omega_1 t) + \frac{\{1-e^{-2[\omega_1 t/(n-3/2)-1]}\}}{\pi(n-1/2)}H[\omega_1 t-(n-3/2)]\right]. \tag{B8}$$

Fig. B1b shows the plots of displacement $u_{n,1}$ for $n=11$, calculated by formulas (B5) (thin curve) and (B8) (thick curve). We see that the thick and thin curves practically coincide. In Fig. B1a and b, the horizontal dashed lines correspond to the static value of displacement calculated by (B6). The vertical dashed lines correspond to the arrival time of longitudinal wave propagating with velocity $c_1$, i.e., to $t = (n-1/2)/c_1$. Fig. B1b shows that solution (B8), which describes a longitudinal wave propagating with velocity $c_1$, oscillates around the static value of displacement determined by (B6).

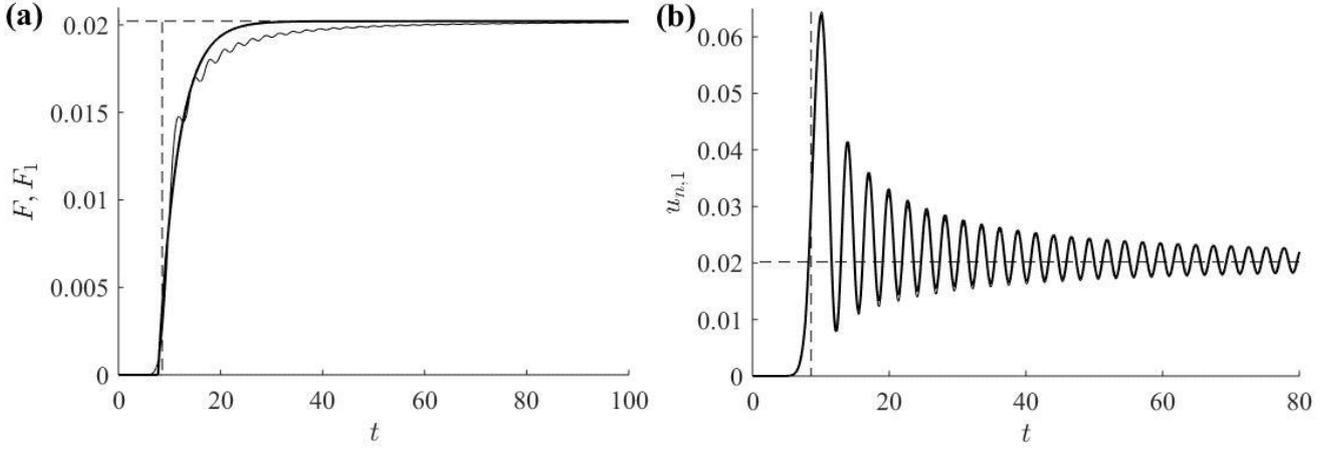

**Fig. B1.** (a) Plots of functions $F(n, \omega_1 t)$ (thin curve) and $F_1(n, \omega_1 t)$ (thick curve) versus time for $n=11$. (b) Plots of displacement $u_{n,1}$ versus time calculated by (B5) (thin curve) and by (B8) (thick curve) for $n=11$.

Using (A4) and (A7), we obtain from (B2), (B3), (B8) as $t \to \infty$ (or, which is the same, $n \to \infty$):

$$u_{n,1}(t) \sim \frac{1}{\omega_1^2} \left[ \frac{\mathrm{Ai}(\kappa_{n-1,1})\mathrm{Ai}(\kappa_{n,1})}{(\omega_1 t/2)^{2/3}} + \frac{\{1 - e^{-2[\omega_1 t/(n-3/2)-1]}\}}{\pi(n-1/2)} H[\omega_1 t - (n-3/2)] \right], \qquad (B9)$$

$$\dot{u}_{n,1}(t) \sim \frac{\mathrm{Ai}^2(\kappa_{n-1,1}) - \mathrm{Ai}^2(\kappa_{n,1})}{\omega_1 (\omega_1 t/2)^{2/3}}, \qquad (B10)$$

$$\ddot{u}_{n,1}(t) \sim \frac{4}{\omega_1 t} \left[ \mathrm{Ai}(\kappa_{n,1})\mathrm{Ai}'(\kappa_{n,1}) - \mathrm{Ai}(\kappa_{n-1,1})\mathrm{Ai}'(\kappa_{n-1,1}) \right]. \qquad (B11)$$